\documentclass[a4paper,11pt]{article}
\pdfoutput=1 

\usepackage{jheppub} 
\usepackage[vcentermath]{youngtab}
\usepackage{url,amsmath,amssymb,amsfonts,amsxtra, mathrsfs,graphics,graphicx,amsthm,epsfig, youngtab,bm,longtable,float,tikz,caption}
\usepackage[T1]{fontenc} 
\usepackage[margin=0.5cm]{caption}
\usepackage{jheppub}
\usepackage{float}
\restylefloat{table}
\usepackage{comment}

\newcommand {\apgt} {\ {\raise-.5ex\hbox{$\buildrel>\over\sim$}}\ }
\newcommand {\aplt} {\ {\raise-.5ex\hbox{$\buildrel<\over\sim$}}\ }

\makeatletter\@addtoreset{equation}{section}\makeatother

\makeatletter
\newcommand{\ellSN}{\mathop{\operator@font sn}\nolimits}
\newcommand{\ellCN}{\mathop{\operator@font cn}\nolimits}
\newcommand{\ellDN}{\mathop{\operator@font dn}\nolimits}
\newcommand{\ellAM}{\mathop{\operator@font am}\nolimits}
\newcommand{\ellK}{\mathop{\smash{\operator@font K}\vphantom{a}}\nolimits}
\newcommand{\ellE}{\mathop{\smash{\operator@font E}\vphantom{a}}\nolimits}
\makeatother

\makeatletter
\def\mr@ignsp#1 {\ifx\:#1\@empty\else #1\expandafter\mr@ignsp\fi}%
\newcommand{\multiref}[1]{\begingroup
\xdef\mr@no@sparg{\expandafter\mr@ignsp#1 \: }%
\def\mr@comma{}%
\@for\mr@refs:=\mr@no@sparg\do{\mr@comma\def\mr@comma{,}\ref{\mr@refs}}%
\endgroup}
\makeatother

\newcommand{\hypref}[2]{\ifx\href\asklfhas #2\else\href{#1}{#2}\fi}

\renewcommand{\eqref}[1]{(\multiref{#1})}

\title{Bethe/Gauge correspondence in odd dimension: modular double, non-perturbative corrections and open topological strings}

\author[a]{Antonio Sciarappa}

\affiliation[a]{School of Physics, Korea Institute for Advanced Study, \\
Seoul 130-722, Korea}

\emailAdd{asciara@kias.re.kr}

\abstract{Bethe/Gauge correspondence as it is usually stated is ill-defined in five dimensions and needs a ``non-perturbative'' completion; a related problem also appears in three dimensions. It has been suggested that this problem, probably due to incompleteness of Omega background regularization in odd dimension, may be solved if we consider gauge theory on compact $S^5$ and $S^3$ geometries. We will develop this idea further by giving a full Bethe/Gauge correspondence dictionary on $S^5$ and $S^3$ focussing mainly on the eigenfunctions of (open and closed) relativistic 2-particle Toda chain and its quantized spectral curve: these are most properly written in terms of non-perturbatively completed NS open topological strings. A key ingredient is Faddeev's modular double structure which is naturally implemented by the $S^5$ and $S^3$ geometries.}

\preprint{KIAS-P16039}

\begin{document}

\maketitle

\newpage

\section{Introduction}

The analysis of the Coulomb branch moduli space of vacua of four dimensional $\mathcal{N}=2$ $SU(2)$ theories carried by Seiberg and Witten \cite{Seiberg:1994rs,Seiberg:1994aj}\footnote{See also \cite{Cecotti:1990fq} for earlier work on the subject.} led to a huge number of new developments in the study and understanding of supersymmetric gauge theories.
One of the most important consequences of this analysis is that it provides a clear way to relate the Coulomb branch of four dimensional $\mathcal{N}=2$ theories to complex algebraic integrable system \cite{Gorsky:1995zq,Donagi:1995cf,Martinec:1995by}; in particular, in the pure $SU(N)$ case the Seiberg-Witten curve can be shown to coincide with the spectral curve of the classical $N$-particle Toda chain, while for a generic theory of class $\mathcal{S}$ \cite{Gaiotto:2009we} the associate system is a Hitchin system \cite{Gaiotto:2009hg}. This remains true, with the appropriate modifications, if one considers $\mathcal{N}=1$ five dimensional gauge theories \cite{Nekrasov:1996cz}: focussing again on pure $SU(N)$, its five dimensional Seiberg-Witten curve is the same as the spectral curve of the ``relativistic'' version of the Toda chain \cite{ruijsenaars1990,Ruij}. \\
Subsequent developments \cite{2002hep.th....6161N,2003hep.th....6238N} based on the study of four and five dimensional theories in the presence of Omega background made it possible to extend the correspondence between supersymmetric gauge theories and integrable systems 
to the situation in which the integrable system gets quantized (what is nowadays known as Bethe/Gauge correspondence \cite{2010maph.conf..265N}). Letting $\epsilon_1$, $\epsilon_2$ be the parameters of Omega background, quantization of the integrable system requires considering the so-called Nekrasov-Shatashvili (NS) limit on the gauge theory side in which $\epsilon_2 \rightarrow 0$ while $\epsilon_1$ remains finite (or viceversa); the remaining parameter $\epsilon_1$ will play the role of the Planck constant $\hbar$. In a natural way, turning off the Omega background completely (i.e. sending also $\epsilon_1 \propto \hbar \rightarrow 0$) brings us back to the Seiberg-Witten / classical integrable system correspondence discussed above. \\
Bethe/Gauge correspondence in four and five dimensions is a fascinating subject still under development (see for example \cite{Nvideo} for an overview of the current state of the art). In the four dimensional case ($\mathbb{R}^2_{\epsilon_1} \times \mathbb{R}^2$) many pieces of evidence for the validity of this correspondence have been collected, in particular numerical evaluation of the spectrum of quantum mechanical systems can be shown to agree with the results obtained via gauge theory, at least for quantum mechanical problems with single vacuum potentials \cite{2015arXiv151102860H}; in the case of periodic potentials such as Mathieu and Lam\'{e} systems the story is more complicated and may also require the use of resurgence techniques \cite{Krefl:2014nfa,Piatek:2014lma,Piatek:2015jva,2015JHEP...02..160B,2016arXiv160304924D}. In five dimensions instead ($\mathbb{R}^2_{\epsilon_1} \times \mathbb{R}^2 \times S^1$) it has been recently pointed out in \cite{2013arXiv1308.6485K,Grassi:2014zfa,2015arXiv151102860H} that Bethe/Gauge correspondence, as it is stated, is incomplete: for example the numerical spectrum does not coincide with the gauge theory results, and moreover gauge theory quantities in the NS limit are ill-defined for values of the Planck constant (i.e. $\epsilon_1$) which should actually be perfectly admissible. Nevertheless the correspondence can be restored, and these problems solved, by properly considering the contribution of quantum mechanical instantons, non-perturbative in $\hbar$; on the gauge theory side this translates into finding a ``non-perturbative completion'' of the five dimensional partition function in the NS limit, or of the refined closed topological string in the NS limit in geometric engineering language, as properly done in \cite{Grassi:2014zfa,2015arXiv151102860H}. \\
Another non-perturbative completion of refined topological strings, in principle different from the one of \cite{Grassi:2014zfa,2015arXiv151102860H}, has been proposed in \cite{2012arXiv1210.5909L}; the two proposals can been shown to be compatible \cite{2015arXiv150704799H} if one takes into consideration the observations in \cite{Wang:2015wdy}. The idea of \cite{2012arXiv1210.5909L} involves considering the partition function $\mathcal{Z}_{S^5}$ of the gauge theory on the squashed $S^5$ \cite{Kim:2012ava,Kim:2012qf,Kallen:2012cs,Hosomichi:2012ek,Kallen:2012va,Imamura:2012xg,Imamura:2012bm}: thinking of $S^5$ as an $(S^1)^3$ fibration over a triangle, the integrand of $\mathcal{Z}_{S^5}$ is expected to factorize into three copies of the flat space partition function $\mathcal{Z}_{\mathbb{R}^4 \times S^1}$, each copy corresponding to one of the vertices of the triangle
\begin{equation}
\mathcal{Z}_{S^5} \sim \int [da] \, \mathcal{Z}_{\mathbb{R}^4 \times S^1}^{(1)} \mathcal{Z}_{\mathbb{R}^4 \times S^1}^{(2)} \mathcal{Z}_{\mathbb{R}^4 \times S^1}^{(3)}
\end{equation}
and the proposal is that $\mathcal{Z}_{\mathbb{R}^4 \times S^1}^{(2)} \mathcal{Z}_{\mathbb{R}^4 \times S^1}^{(3)}$ provides the desired non-perturbative completion of $\mathcal{Z}_{\mathbb{R}^4 \times S^1}^{(1)}$. In the following we will be interested in the ``NS limit'' of this geometry\footnote{This is not what has been considered in \cite{Grassi:2014zfa,2015arXiv151102860H} and \cite{2012arXiv1210.5909L}: there the authors were looking for a non-perturbative completion of unrefined topological strings, while here we are doing something similar but for NS topological strings by following a suggestion in \cite{2012arXiv1210.5909L}.}: in this limit one of the three copies (say $\mathcal{Z}_{\mathbb{R}^4 \times S^1}^{(3)}$) drastically simplifies, and we remain with the statement that $\mathcal{Z}_{\mathbb{R}^4 \times S^1}^{(2), NS}$ is the non-perturbative completion of $\mathcal{Z}_{\mathbb{R}^4 \times S^1}^{(1), NS}$. 
Following this line of reasoning while keeping in mind the lesson of \cite{Grassi:2014zfa,2015arXiv151102860H} one can derive non-perturbatively corrected quantization conditions for the associated quantum integrable system which are more refined than the ones one would get by only considering $\mathcal{Z}_{\mathbb{R}^4 \times S^1}^{(1), NS}$ as originally suggested in \cite{2010maph.conf..265N} and basically correspond to the ones given in \cite{Wang:2015wdy}. \\
In this paper we will develop the idea of \cite{2012arXiv1210.5909L} further by analysing a class of codimension two and four defects for the $\mathcal{N}=1$ $SU(N)$ theory on $S^5$; these defects will wrap one of the three $S^3$ or $S^1$ respectively. In the NS limit only one $S^3$ and two $S^1$ survive, and our defects wrapping these submanifolds have a natural interpretation from the quantum integrable system point of view: they are eigenfunctions and eigenvalues of the commuting set of quantum operators defining the relativistic $N$-particle Toda chain as well as its Faddeev's ``modular dual'', related to the Toda chain by the exchange $\hbar \leftrightarrow 1/\hbar$. A careful treatment of the problem inspired by \cite{Grassi:2014zfa,2015arXiv151102860H} provides a non-perturbative completion of the prescription in \cite{2010maph.conf..265N} for computing eigenfunctions and eigenvalues of relativistic Toda: as we will see if one only considers defects in $\mathbb{R}^2_{\epsilon_1} \times \mathbb{R}^2_{\epsilon_2} \times S^1$ in the NS limit the eigenfunctions are ill-defined for some value of $\hbar$ and ambiguous for all values of $\hbar$, while going to $S^5$ solves both problems at once. At the level of integrable systems, going to $S^5$ is equivalent to properly take into account the existence of the modular dual relativistic Toda chain, as done in \cite{Kharchev:2001rs} for the open relativistic Toda chain case (based on previous observations in \cite{1995LMaPh..34..249F,Faddeev:1999fe});
remarkably, the modular double structure of the quantum system allows us to consider self-adjoint operators also for $\hbar$ complex. The importance of the modular double structure has also been very recently remarked in \cite{Nedelin:2016gwu}. \\
An important point has to be stressed. Our analysis will initially correspond to the gauge theory version of the computations carried in \cite{Grassi:2014zfa,2015arXiv151102860H}, which have been performed via closed topological strings in the Calabi-Yau geometry engineering our five dimensional $SU(N)$ theory. However five-dimensional gauge theory quantities may be affected by problems of convergence in the instanton counting parameter $Q_{5d}$, and moreover they are hard to work with if one is interested in analysing their non-perturbative completions; this is why at the end we will have to re-express our results in terms of (open or closed) topological strings, which have better convergence properties and are easier to deal with. Nevertheless the gauge theory setting is also helpful because many quantities (such as the eigenfunctions) are easier to compute in this way and seems more convenient to study finite-difference equations since it involves series in a single parameter; hopefully this work will provide ideas on how to obtain the eigenfunctions of relativistic quantum mechanical systems directly from open topological strings. \\

\noindent The plan of the paper is the following. We start in Section \ref{sec2} by reviewing what is known about relativistic Toda chains and their modular double structure; we then move to discuss Bethe/Gauge correspondence on flat space and argue that in odd dimension Bethe/Gauge correspondence is only well-defined on compact spaces. Going to compact spaces automatically implements the modular double structure of the underlying relativistic quantum mechanical system. 
\\
Section \ref{sec3} focusses on our main example, the quantum relativistic 2-particle open and closed Toda chain. While the known solution for the open Toda chain \cite{Kharchev:2001rs} admits a natural interpretation as a gauge theory on the squashed three-sphere, the closed chain requires to uplift the discussion to a particular limit of the squashed five-sphere. We will describe what gauge theory can say about non-perturbative completion of the eigenfunctions of the closed Toda Hamiltonians and of the operator associated to the quantization of its spectral curve, and comment on Fourier transformation of these eigenfunctions. At the end of the Section we rewrite our eigenfunctions in terms of open refined topological strings in the NS limit in order to properly analyse their non-perturbative completion. \\
Section \ref{sec4} contains a final discussion and various comments. \\

\noindent A related work which will consider a different approach to the problem of studying eigenfunctions of relativistic quantum integrable systems is in preparation \cite{Marino:2016rsq}.
 

\section{Review of Toda chains and Bethe/Gauge correspondence} \label{sec2}

In this Section we review what is known about relativistic $N$-particle Toda chains (open and closed) from many different point of views: quantum integrable systems, supersymmetric gauge theories, topological strings. The goal of the Section is mainly to fix notations and contextualize the problem we want to solve. 

\subsection{Relativistic open and closed Toda chains} \label{subs2.1}

The quantum relativistic\footnote{See \cite{Braden:1997nc} for more details on the proper interpretation of the word ``relativistic''.} $N$-particle Toda chain (open and closed) was introduced by Ruijsenaars in \cite{ruijsenaars1990}. 
This is simply a quantum mechanical system of $N$ particles on a line with an interaction defined by the Hamiltonian\footnote{We are fixing the interaction constant to one for simplicity.}
\begin{equation}
\widehat{H}_1 = \sum_{n=1}^N \left[ 1 + q^{-1/2} e^{\frac{2\pi}{\omega_2} (x_n - x_{n+1})} \right] e^{\omega_1 p_n} \label{clT}
\end{equation}
with boundary condition $x_{N+1} = x_1 - \ln Q$ imposed. Here $x_n$, $p_n$ are coordinate and momentum operators of the $N$ particles and satisfy the commutation relations $[x_m, p_n] = i \delta_{m,n}$, while combinations of $\omega_1$, $\omega_2$ $\in$ $\mathbb{R}_+$ are related to the Planck constant $\hbar$ and the ``speed of light'' $c$
; finally, we defined $q = e^{2\pi i \omega_1/\omega_2}$. The parameter $Q = e^{-8\pi^2/(g^2 \omega_2)} \in \mathbb{R}_+$ has to be thought as a useful accessory parameter which allows us to distinguish between the open chain ($Q = 0$) and the closed chain ($Q = 1$); as we will see, gauge theoretical computations for the closed chain typically involve power series expansions in $Q$, so we prefer to keep this additional parameter in the definition of the Hamiltonian. \\
An important point we want to stress is that there is a big qualitative difference between the open Toda chain and the closed one. In the open case $Q=0$ the potential of the theory does not have a stable vacuum (it actually has a runaway direction), therefore the quantum theory will have a continuous spectrum and non-normalizable wave-functions, exactly as for the case of a free particle; on the other hand, as soon as $Q > 0$ the potential will acquire a stable vacuum, which implies a discrete set of energies at the quantum level and $L^2$-normalizable wave-functions. \\
Because of its integrability the Toda system admits $N-1$ additional commuting operators $\widehat{H}_k$, $k = 2, \ldots, N$: for example
\begin{equation}
\begin{split}
\widehat{H}_N & = \sum_{n=1}^N e^{\omega_1 p_n} \\
\widehat{H}_{N-1} & = \widehat{H}_N \sum_{n=1}^N \left[ 1 + q^{-1/2} e^{\frac{2\pi}{\omega_2} (x_n - x_{n+1})} \right] e^{-\omega_1 p_{n+1}} 
\end{split}
\end{equation}
with $p_{N+1} = p_1$. Imposing $\widehat{H}_N = 1$ is equivalent to decouple the center of mass of the system. In the context of Baxter $\mathcal{T}$-$\mathcal{Q}$ relations and Quantum Inverse Scattering Method \cite{BAXTER1972193,BAXTER19731,Bax} it is often useful to collect all the quantum operators $\widehat{H}_k$, $k = 1, \ldots, N$ into a generating function $\mathbf{t}_N(w)$ (the transfer matrix)
\begin{equation}
\mathbf{t}_N (w) = \sum_{k=0}^N (-1)^k w^{\frac{N}{2} - k} \widehat{H}_k 
\;\;\;,\;\;\; \widehat{H}_0 = 1
\end{equation}
which is nothing but the trace of the monodromy matrix, constructed from the quantum Lax operator; for the case at hand these operators can be found for example in \cite{KT}. Commutativity of the $\widehat{H}_k$ is ensured by the condition
\begin{equation}
[\mathbf{t}_N (w), \mathbf{t}_N (z)] = 0
\end{equation} 
If we decouple the center of mass of the system, solving the closed chain quantum problem means finding discrete eigenvalues $\vec{E} = (E_1, \ldots, E_{N-1})$ and $L^2(\mathbb{R}^{N-1})$-normalizable common eigenfunctions $\psi_{\vec{E}}(\vec{x})$ of the $N-1$ quantum operators $\widehat{H}_k$, $k=1, \ldots, N-1$ such that
\begin{equation}
\mathbf{t}_N (w) \psi_{\vec{E}}(\vec{x}) = t_N (w) \psi_{\vec{E}}(\vec{x})
\end{equation}
with
\begin{equation}
t_N (w) = \sum_{k=0}^N (-1)^k w^{\frac{N}{2} - k} E_k \;\;\;,\;\;\; E_0 = E_N = 1 \label{gfe}
\end{equation}
generating function of the eigenvalues. For the open chain the problem is similar, but the spectrum is continuous and the eigenfunctions are not $L^2$-normalizable. \\
The description of the classical relativistic Toda system as an algebraic integrable system requires the introduction of the spectral curve for the model; this is a Riemann surface embedded in $(w,y) \in \mathbb{C^*} \times \mathbb{C^*}$ given by
\begin{equation}
y + Q\, y^{-1} = t_N (w) \label{spe}
\end{equation}
which allows us to compute the action variables as the periods of an appropriate differential. The spectral curve admits many different representations, related by change of coordinates; for example a slightly different realization (Baxter-like) is given by 
\begin{equation}
(i)^{-N} y + Q\, (i)^N y^{-1} = t_N (w) \label{spe2}
\end{equation}
while another realization (toric CY-like) is
\begin{equation}
y + Q\, w^{N} y^{-1} = w^{\frac{N}{2}} t_N (w) 
\end{equation}
As we will see in the following, it is from this spectral curve that we can see the connection among relativistic $N$-particle Toda, five-dimensional supersymmetric theories and topological strings. At the quantum level the spectral curve gets promoted to a finite-difference equation: by redefining $w=e^x$, $y = e^p$ and quantizing the $(x,p)$ space via $[x,p] = i \hbar$, equation \eqref{spe2} reduces to
\begin{equation}
(i)^{-N} \mathcal{Q}(x - i \hbar) + Q\, (i)^{N} \mathcal{Q}(x + i \hbar) = t_N(w) \mathcal{Q}(x) \label{quantumspec}
\end{equation} 
which is known as Baxter $\mathcal{T}$-$\mathcal{Q}$ equation, a central object in the study of the solution to quantum mechanical problems in the framework of Quantum Inverse Scattering.


\subsection{Modular duality} \label{subs2.2}

An interesting observation put forward in \cite{Kharchev:2001rs} (motivated by earlier works \cite{1995LMaPh..34..249F,Faddeev:1999fe}) is the existence of a \textit{modular dual} relativistic Toda system, which is not independent from the one we considered above since they are related by the exchange $\omega_1 \leftrightarrow \omega_2$; that is, there exists another set of $N$ commuting quantum operators $\widehat{\widetilde{H}}_k$, the first one being
\begin{equation}
\widehat{\widetilde{H}}_1 = \sum_{n=1}^N \left[ 1 + \widetilde{q}^{-1/2} e^{\frac{2\pi}{\omega_1} (x_n - x_{n+1})} \right] e^{\omega_2 p_n} \;\;\;, \;\;\; \widetilde{q} = e^{2 \pi i \omega_2/\omega_1}
\end{equation}
which also commute with the operators $\widehat{H}_k$ of the original system. The boundary condition in this case is $x_{N+1} = x_1 - \ln \widetilde{Q}$ with $\widetilde{Q} = e^{-8\pi^2/(g^2 \omega_1)} \in \mathbb{R}_+$. By constructing the dual transfer matrix
\begin{equation}
\widetilde{\mathbf{t}}_N (\widetilde{w}) = \sum_{k=0}^N (-1)^k \widetilde{w}^{\frac{N}{2} - k} \widehat{\widetilde{H}}_k 
\;\;\;,\;\;\; \widehat{\widetilde{H}}_0 = 1
\end{equation}
commutativity of the two sets of Toda operators $\widehat{H}_k$, $\widehat{\widetilde{H}}_k$ is encoded in the relations
\begin{equation}
\begin{split}
[\,\mathbf{t}_N (w), \mathbf{t}_N (z) \,] & = 0 \\
[\,\widetilde{\mathbf{t}}_N (\widetilde{w}), \widetilde{\mathbf{t}}_N (\widetilde{z}) \,] & = 0 \\
[\,\mathbf{t}_N (w), \widetilde{\mathbf{t}}_N (\widetilde{w})\,] & = 0 
\end{split}
\end{equation}
We will have a spectral curve also for the classical dual system, of the form
\begin{equation}
\widetilde{y} + \widetilde{Q}\, \widetilde{y}^{-1} = \widetilde{t}_N (\widetilde{w}) \label{sped}
\end{equation}
in the realization \eqref{spe}. The appearance of the dual system is not something one can ignore since it origins from representation theory. It has long been known that the spectral problem of the non-relativistic Toda chain can be reduced to considering the representation theory of semisimple Lie groups \cite{Konstant}; it is therefore natural to extend this approach to the relativistic Toda system by studying the representation theory of quantum groups $U_q(\mathfrak{g})$ based on the algebras $\mathfrak{g} = \mathfrak{sl}(N)$ or $\mathfrak{gl}(N)$. The analysis carried out in \cite{Kharchev:2001rs} shows that the correct treatment of the problem actually requires to consider the representation theory of the \textit{modular double} $U_q(\mathfrak{g}) \otimes U_{\widetilde{q}}({}^{L}\mathfrak{g})$ with ${}^{L}\mathfrak{g}$ Langlands dual of $\mathfrak{g}$. At the practical level, among other things this allowed the authors of \cite{Kharchev:2001rs} to construct unambiguous eigenfunctions of the relativistic Toda system (aka Whittaker vectors), at least for the open chain. In fact, differently from non-relativistic quantum mechanical systems in which the eigenfunctions are only defined modulo a constant, eigenfunctions of the Toda operators $\widehat{H}_k$ are only defined modulo a function of period $i \omega_1$ (i.e. a quasi-constant) which are unaffected by operators like $e^{-i \omega_1 \partial_x}$; this ambiguity can be reduced to the usual overall constant normalization by requiring them to simultaneously be eigenfunctions of the $\widehat{\widetilde{H}}_k$'s as well. We will have to keep this in mind when we construct the eigenfunctions of the closed Toda chain with gauge theory techniques. \\
As remarked in \cite{Kharchev:2001rs}, the existence of the modular dual Hamiltonians allows us to make sense of the quantum mechanical problem also for $\omega_1$, $\omega_2$ complex: in fact the Toda operators $\widehat{H}_k$, $\widehat{\widetilde{H}}_k$ are self-adjoint on $L^2(\mathbb{R}^{N-1})$ when $\omega_1$, $\omega_2$ are real, while the combinations $\widehat{H}_k + \widehat{\widetilde{H}}_k$ and $i(\widehat{H}_k - \widehat{\widetilde{H}}_k)$ are self-adjoint for $\overline{\omega}_1 = \omega_2$. Only in this sense it is reasonable to analyse the Toda system for $\hbar$ complex, as done for example in \cite{Kashani-Poor:2016edc} and as we will sometimes do in the following.

\subsection{Bethe/Gauge correspondence} \label{subs2.3}

As observed in \cite{Gorsky:1995zq,Donagi:1995cf,Martinec:1995by}, there is a deep connection between classical algebraic integrable systems (such as the non-relativistic open and closed Toda chains) and four-dimensional $\mathcal{N}=2$ theories on $\mathbb{R}^4$: this is most easily seen by comparing the Seiberg-Witten curve of the 4d theory with the spectral curve of the corresponding non-relativistic classical integrable system. \\
This connection can be extended to include five-dimensional theories and classical relativistic integrable systems \cite{Nekrasov:1996cz}: in particular, the classical closed relativistic $N$-particle Toda chain (with center of mass factored out) is associated to the 5d $\mathcal{N}=1$ $SU(N)$ theory on $\mathbb{R}^4 \times S^1_R$. In fact one can compare
the Seiberg-Witten curve of the 5d $SU(N)$ theory \cite{Nekrasov:1996cz,Kim:2014nqa} with the spectral curve of the $N$-particles Toda chain \cite{Ruij} and check that they coincide. For example, for $N=2$ the Seiberg-Witten curve reads
\begin{equation}
y + Q_{5d}\, y^{-1} = w - U + w^{-1} \label{SW2}
\end{equation}
with $Q_{5d} = e^{-8\pi^2 R/g^2_{5d}}$, while the 2-particles Toda spectral curve is
\begin{equation}
y + Q\, y^{-1} = w - E_1 + w^{-1}  \label{sc2}
\end{equation}
with $E_1$ classical energy of the system. When the 5d Yang-Mills coupling $g_{5d}$ is turned off, i.e. $Q_{5d} \rightarrow 0$, our Seiberg-Witten curve \eqref{SW2} reduces to the spectral curve for the open Toda chain. In this correspondence the energy $E_1$ of the classical Toda Hamiltonian $H_1$ is mapped to the vacuum expectation value $U$ of the trace in the fundamental representation of a 5d $SU(2)$ Wilson loop wrapping the $S^1_R$ circle. \\

One may expect this correspondence could be promoted to the quantum level, were we able to introduce a parameter playing the role of $\hbar$ on the gauge theory side. In the case of four-dimensional theories, the proposal (named Bethe/Gauge correspondence) put forward in \cite{2010maph.conf..265N} is the following: we can start by considering the 4d theory on $\mathbb{R}^2_{\epsilon_1} \times \mathbb{R}^2_{\epsilon_2}$ in the presence of the most general Omega background and compute its partition function \cite{2002hep.th....6161N,2003hep.th....6238N}
\begin{equation}
Z_{4d}(\epsilon_1, \epsilon_2, Q_{4d}, \vec{a}) \,=\, \text{exp} \left[ \mathcal{F}_{4d}(\epsilon_1, \epsilon_2, Q_{4d}, \vec{a}) \right]
\end{equation} 
with $\vec{a}$ vacuum expectation values of the scalar fields in the Cartan part of the $\mathcal{N}=2$ vector multiplet. 
We can now take the so-called Nekrasov-Shatashvili limit $\epsilon_2 \rightarrow 0$, $\epsilon_1$ fixed and construct the function
\begin{equation}
\mathcal{W}_{4d}(\epsilon_1, Q_{4d}, \vec{a}) = \lim_{\epsilon_2 \rightarrow 0} \left[ \epsilon_2 \,\mathcal{F}_{4d}(\epsilon_1, \epsilon_2, Q_{4d}, \vec{a}) \right]
\end{equation}
which is known as \textit{effective twisted superpotential}. This function can then be interpreted as the Yang-Yang function for the quantized version of the classical integrable system which was associated to our gauge theory in absence of Omega background; $\epsilon_1$ naturally corresponds to the Planck constant $\hbar$.
Therefore the equations determining the supersymmetric vacua in the Coulomb branch\footnote{As mentioned in \cite{2010maph.conf..265N} quantization of $a_D$ is related to a quantum mechanical problem with $L^2(\mathbb{R})-$normalizable eigenfunctions such as our 2-particles Toda system where the potential is basically a hyperbolic cosine, while quantization of $a$ will lead to a problem with quasi-periodic eigenfunctions such as a system with a cosine potential. The second case is extremely more involved: see \cite{Krefl:2014nfa,Piatek:2014lma,Piatek:2015jva,2015JHEP...02..160B,2016arXiv160304924D} for a discussion on the spectral problem of the Mathieu equation, related to four-dimensional $\mathcal{N} = 2$ $SU(2)$ theory.}
\begin{equation}
a^D_i = \dfrac{\partial \mathcal{W}_{4d}}{\partial a_i} = 2\pi i \left( n_i + \dfrac{1}{2} \right) \;\;\;,\;\;\; n_i \in \mathbb{Z} \label{BAE4d}
\end{equation}
coincide with the Bethe Ansatz Equations for the quantum integrable system: these fix the values of $a_i$ (which are parameters entering in the Bethe Ansatz for the eigenfunctions of the system) in such a way to ensure $L^2(\mathbb{R}^{N-1})-$normalizability and single-valuedness of the eigenfunctions and determine the discrete spectrum of the system. On the gauge theory side eigenvalues and eigenfunctions are given by the vacuum expectation value of observables associated to appropriate codimension four and two defects respectively, evaluated at the vacua determined by \eqref{BAE4d}, while the Hamiltonians of the integrable system coincide with the twisted chiral ring operators of the effective two-dimensional theory living on the decompactified $\mathbb{R}^2$. In the limit $Q_{4d} \rightarrow 0$ the theory has a moduli space of supersymmetric vacua instead of a finite number of vacua, equations \eqref{BAE4d} degenerate, and therefore the parameters $a_i$ remain continuous: this corresponds to having a quantum mechanical system whose potential has not a proper single vacuum, which implies a continuous spectrum and non $L^2$-normalizable eigenfunctions. \\

\begin{figure}[h]
\centering
\includegraphics[width=0.75\textwidth]{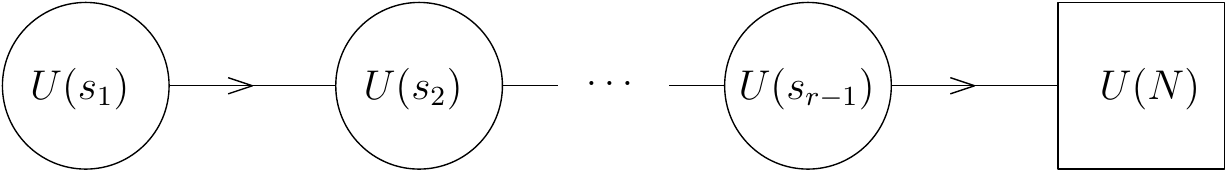}
\caption{3d monodromy defect theory for 5d $\mathcal{N}=1$ $U(N)$ at generic $\rho$.}
\label{Fig:figgeneral}
\end{figure}

\begin{figure}[h]
\centering
\includegraphics[width=1\textwidth]{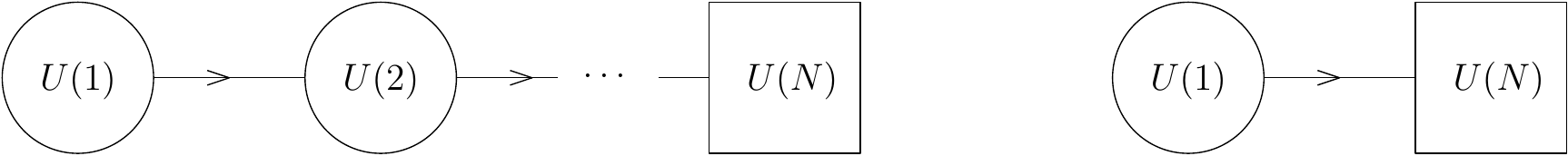}
\caption{Full ($\rho = (1^N)$, left) and simple ($\rho = (1,N-1)$, right) monodromy defect.}
\label{Fig:figfullsimple}
\end{figure}

As far as five dimensional theories are concerned, the situation is less clear. The proposal in \cite{2010maph.conf..265N} is, strictly speaking, only formulated for four dimensional gauge theories; still one can imagine that a similar story might also be valid in five dimensions. Let us follow this idea for the time being, and specialize to 5d $\mathcal{N}=1$ $SU(N)$ theories on Omega background $\mathbb{R}^2_{\epsilon_1} \times \mathbb{R}^2_{\epsilon_2} \times S^1_R$: from the discussion above, we expect the NS limit of this theory to be related to the quantum relativistic $N$-particle closed Toda chain we introduced earlier in this Section. From the partition function 
\begin{equation}
Z_{5d}(\epsilon_1, \epsilon_2, R, Q_{5d}, \vec{a}) = \text{exp} \left[ \mathcal{F}_{5d}(\epsilon_1, \epsilon_2, R, Q_{5d}, \vec{a}) \right]
\end{equation} 
we can recover the relativistic Toda Yang-Yang function $\mathcal{W}_{5d}$ by taking the Nekrasov-Shatashvili limit 
\begin{equation}
\mathcal{W}_{5d}(\epsilon_1, R, Q_{5d}, \vec{a}) = \lim_{\epsilon_2 \rightarrow 0} \left[ \epsilon_2 \, \mathcal{F}_{5d}(\epsilon_1, \epsilon_2, R, Q_{5d}, \vec{a}) \right] \label{YY5d}
\end{equation}
The supersymmetric vacua equations
\begin{equation}
a^D_i = \dfrac{\partial \mathcal{W}_{5d}}{\partial a_i} = 2 \pi i \left( n_i + \dfrac{1}{2} \right) \;\;\;,\;\;\; n_i \in \mathbb{Z} \label{BAE}
\end{equation}
will fix the values of $a_i$ in such a way to ensure single-valuedness of the wave-function. Eigenfunctions and eigenvalues of the relativistic Toda Hamiltonians $\widehat{H}_k$, $k = 1, \ldots, N-1$ have a precise description in terms of $\mathcal{N}=1$ $SU(N)$ gauge theoretical quantities \cite{Nvideo,Gaiotto:2014ina,Bullimore:2014awa}. The eigenvalue $E^{(k)}_{\vec{n}}$ of the $k$-th Hamiltonian is given by the NS limit of the vacuum expectation value of an $SU(N)$ gauge Wilson loop in the $k$-th antisymmetric representation wrapping $S^1_R$ (a codimension four defect), evaluated at the solution of \eqref{BAE} labelled by the set of integers $\vec{n}$: 
\begin{equation}
E^{(k)}_{\vec{n}} = \langle W^{SU(N)}_{\Lambda^k} \rangle^{NS}\Big\vert_{\vec{n}}
\end{equation}
The simultaneous eigenfunctions $\psi_{\vec{n}}(x_1, \ldots, x_N)$ of the Toda Hamiltonians correspond instead to the NS limit of the 5d partition function $Z_{5d,(\rho)}$ in the presence of a particular class of codimension two defects on $\mathbb{R}^2_{\epsilon_1} \times S^1_R \subset \mathbb{R}^2_{\epsilon_1} \times \mathbb{R}^2_{\epsilon_2} \times S^1_R$ known as Gukov-Witten (full) monodromy defects \cite{2006hep.th...12073G,2008arXiv0804.1561G}, again evaluated at the vacuum labelled by $\vec{n}$. In this case it is better to consider the $U(N)$ theory first and later decouple the $U(1)$ factor. The possible $U(N)$ monodromy defects are in one to one correspondence with partitions $\rho = (N_1, N_2, \ldots, N_r)$ of $N$ where $0 \leqslant N_1 \leqslant N_2 \leqslant \ldots \leqslant N_r$ and 
$\sum_{j=1}^r N_j = N$; the partition determines the Levi subgroup $\mathbb{L} = U(N_1) \times U(N_2) \times \ldots \times U(N_r)$ of $U(N)$ left unbroken by the defect. Given $\mathbb{L}$, there is an additional label $\sigma$ which specifies the $N_{\rho} = N!/(N_1!N_2! \ldots N_r!)$ non-equivalent choices of embedding $\mathbb{L}$ into $U(N)$. At fixed $\sigma$ the monodromy defect corresponds to prescribe a singular behaviour for the gauge field 
\begin{equation}
\oint_{\vert z_2 \vert = \delta} A^a = 2\pi m^a \;\;\;,\;\;\; a = 1, \ldots, N
\end{equation}
in the complex plane orthogonal to the defect; here 
\begin{equation}
m^a = (\underbrace{m_1, \ldots, m_1}_{N_1}, \underbrace{m_2, \ldots, m_2}_{N_2},\ldots, \underbrace{m_r, \ldots, m_r}_{N_r}) \label{mono}
\end{equation}
In the case of a full\footnote{See \cite{Bullimore:2014awa} for a discussion on the meaning of defects other than full and simple in terms of eigenfunctions of the associated quantum integrable system.} monodromy defect $\rho = (1^N)$ the parameters $(m_1, m_2, \ldots, m_N)$ correspond to the coordinates $(x_1, x_2, \ldots, x_N)$ of the Toda Hamiltonian \eqref{clT} and we have 
\begin{equation}
\psi_{\vec{n}}(x_1, \ldots, x_N) \,=\, \langle Z_{5d,(1^N)}(x_1, \ldots, x_N) \rangle^{NS} \Big\vert_{\vec{n}}
\end{equation}
Monodromy defects can alternatively be described as coupling the 5d theory to a 3d $\mathcal{N}=2$ theory living on the defect \cite{2009AdTMP..13..721G}; the relevant quiver 3d theory for a generic partition $\rho$ is represented in Figure \ref{Fig:figgeneral} and contains a series of $U(s_i)$ gauge groups of rank $s_i = N_1 + \ldots + N_i$, $i = \, \ldots, r-1$ (so that the last, flavour node has rank $s_r = N$). The quivers for the special cases of full ($\rho = (1^N)$) and simple ($\rho = (1,N-1)$) defects are shown in Figure \ref{Fig:figfullsimple}. The full defect 3d theory can be thought of as the chiral limit of the $T[U(N)]$ quiver. The $N_{\rho}$ possible embeddings correspond to the number of supersymmetric vacua of the 3d theory, while the parameters $(m_1, m_2, \ldots, m_r)$ are related to 3d Fayet-Iliopoulos parameters.

The eigenfunctions $\psi_{\vec{n}}(x_1, \ldots, x_N)$ therefore also correspond to the NS limit of the partition function $Z_{3d/5d}$ of our 5d theory coupled to the 3d theory given by the left quiver in Figure \ref{Fig:figfullsimple}, evaluated at the solution $\vec{n}$ of \eqref{BAE}. If we turn off the 5d gauge coupling, i.e. we take the limit $Q_{5d} \rightarrow 0$, we remain with just the partition function $Z_{3d}$ of the 3d defect theory living on $\mathbb{R}^2_{\epsilon_1} \times S^1_R$, which therefore will correspond to the eigenfunction of the open Toda chain; putatively, $Z_{3d}$ will also correspond to the 3d blocks \cite{2012JHEP...04..120P,Beem:2012mb,2013arXiv1303.2626N} that one can extract from the partition function on the squashed 3-sphere $S^3_{\omega_1, \omega_2}$ \cite{2010JHEP...03..089K,Hama:2011ea} or the $S^2 \times S^1$ index \cite{2011JHEP...04..007I,2009NuPhB.821..241K,2011arXiv1106.2484K}. \\

To summarize, if one extends the Bethe/Gauge correspondence of \cite{2010maph.conf..265N} to five dimensional theories on $\mathbb{R}^2_{\epsilon_1} \times \mathbb{R}^2 \times S^1_R$ and relativistic quantum integrable systems, one gets the following dictionary: 

\renewcommand\arraystretch{1.4}
\begin{center}
\begin{table}[h]
\begin{tabular}{|c|c|}
\hline \textbf{relativistic $N$-particle closed Toda} & \textbf{5d $\mathcal{N}=1$ $SU(N)$ theory on $\mathbb{R}^2_{\epsilon_1} \times \mathbb{R}^2 \times S^1_R$} \\ 
\hline (quantized) spectral curve & (quantized) Seiberg-Witten curve \\ 
\hline Bethe Ansatz Equations & supersymmetric vacua equations \\
\hline quantum Hamiltonians $\widehat{H}_k$ & twisted chiral ring operators \\
\hline eigenvalues $E^{(k)}_{\vec{n}}$ & Wilson loops $\langle W^{SU(N)}_{\Lambda^k} \rangle^{NS}\Big\vert_{\vec{n}}$ \\
\hline eigenfunctions $\psi_{\vec{n}}(x_1, \ldots, x_N)$ & full defect $\langle Z_{3d/5d}(x_1, \ldots, x_N) \rangle^{NS} \Big\vert_{\vec{n}}$ \\
\hline coordinates $(x_1, \ldots, x_N)$ & monodromy (FI) parameters $(m_1, \ldots, m_N)$ \\
\hline boundary condition parameter $Q$ & instanton counting parameter $Q_{5d}$ \\
\hline $\omega_2$ & $1/R$ \\
\hline $\omega_1$ & $-i \epsilon_1$ \\
\hline
\end{tabular} \caption{} \label{tab1}
\end{table}
\end{center}
\renewcommand\arraystretch{1} 

\begin{figure}[h]
\centering
\includegraphics[width=0.35\textwidth]{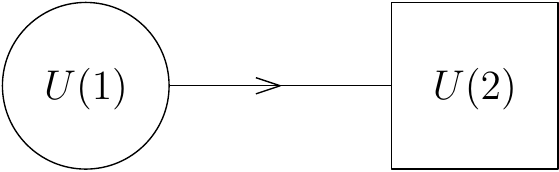}
\caption{Codimension two full (and simple) defect for the 5d $\mathcal{N}=1$ $U(2)$ theory.} \label{Fig:2toda}
\end{figure}

\vspace*{-1 cm}
This proposal has been tested ``off-shell'' (i.e. without imposing \eqref{BAE}) in \cite{Gaiotto:2014ina,Bullimore:2014awa} for the 5d $\mathcal{N} = 1$ $SU(2)$ theory\footnote{In \cite{Bullimore:2014awa}  the 5d $\mathcal{N} = 1^*$ $SU(2)$ theory is also considered: this is related to the $2$-particles elliptic Ruijsenaars-Schneider model, a generalization of the closed relativistic Toda chain.} (see also \cite{Beem:2012mb} for the open case).
In this case the codimension two defect associated to the eigenfunctions corresponds to the chiral limit of the 3d $\mathcal{N}=2^*$ $T[U(2)]$ theory, i.e. the quiver theory in Figure \ref{Fig:2toda}. In these papers the authors explicitly computed $\langle Z_{3d/5d} \rangle^{NS}$ and $\langle W_{\square}^{SU(2)} \rangle^{NS}$ at the first few orders in a series expansion in powers of the instanton counting parameter $Q_{5d}$ and showed that
\begin{equation}
\widehat{H}_1 \langle Z_{3d/5d} \rangle^{NS} = \langle W_{\square}^{SU(2)} \rangle^{NS} \langle Z_{3d/5d} \rangle^{NS}
\end{equation}
Moreover they showed that the 3d blocks of the defect theory, which they extracted from the partition function on $S^3_{\omega_1,\omega_2}$, exactly reproduce the limit $Q_{5d} \rightarrow 0$ of $Z_{3d/5d}$ which we call $Z_{3d}$. 

Nevertheless, this solution is not completely satisfactory: in fact apart from possible problems of convergence as series in $Q_{5d}$, the proposed eigenfunctions are only defined up to quasi-constants (functions periodic in $\hbar$) and only make sense for $\hbar$ complex while they present an infinitely dense set of poles at $\hbar$ real, although $\widehat{H}_1$ is only self-adjoint for real $\hbar$. As noticed in \cite{Grassi:2014zfa,2015arXiv151102860H} (based on \cite{2013arXiv1308.6485K}) in the context of ABJM theory and closed topological strings on local $\mathbb{F}_0$, similar problems also occur when considering the quantization conditions (Bethe Ansatz Equations) for the system: the putative Yang-Yang function one obtains from five-dimensional gauge theories in the NS limit via Bethe/Gauge correspondence \eqref{YY5d} is only defined for $\hbar$ complex and has an infinitely dense set of poles at real $\hbar$, and moreover the spectrum obtained by using \eqref{YY5d} does not match with numerical computation of the eigenvalues\footnote{On the other hand, as showed for example in \cite{2015arXiv151102860H} the Bethe/Gauge correspondence proposal in four dimensions seems to work well: in particular the numerical spectrum of the non-relativistic 2-particle closed Toda chain matches the spectrum one obtains from pure 4d $\mathcal{N}=2$ $SU(2)$.}. In order to cancel these poles and get a spectrum which matches with numerical computations one has to consider contributions coming from quantum-mechanical instantons, i.e. terms of the form $e^{-1/\hbar}$. Although quantum-mechanical instantons are quite unexpected from the form of the Hamiltonian $\widehat{H}_1$, their appearance is less surprising if one remembers that the only consistent relativistic quantum mechanical problem is the one involving also the modular dual Hamiltonian $\widehat{\widetilde{H}}_1$, and that the combined system also makes sense for $\hbar$ complex (in which case one needs to consider the combinations $\widehat{H}_1 + \widehat{\widetilde{H}}_1$ and $i(\widehat{H}_1 - \widehat{\widetilde{H}}_1)$). After properly taking into account the contribution of quantum-mechanical instantons, one finds that the exact quantization conditions are roughly of the form \cite{Grassi:2014zfa,2015arXiv151102860H}
\begin{equation}
a^D_i = \dfrac{\partial}{\partial a_i} \left(\mathcal{W}_{5d}^{\text{WKB}} + \mathcal{W}_{5d}^{\text{np}} \right) = 2 \pi i \left( n_i + \dfrac{1}{2} \right) \;\;\;,\;\;\; n_i \in \mathbb{Z} \label{BAEexact}
\end{equation}
The WKB part is nothing else but the old proposal \eqref{BAE}, while the non-perturbative in $\hbar$ part (np) is a function which only involves contributions of the form $e^{-1/\hbar}$. The new expression \eqref{BAEexact} makes sense also for $\hbar$ real, has no poles in $\hbar$, and provides the correct spectrum of the Hamiltonian $\widehat{H}_1$ (as well as a good definition of non-perturbative completion of refined closed topological strings in the NS limit). The original expression for $\mathcal{W}_{5d}^{\text{np}}$ given in \cite{Grassi:2014zfa,2015arXiv151102860H} was actually quite involved; the important observation of \cite{Wang:2015wdy} simplifies it by showing that $\mathcal{W}_{5d}^{\text{WKB}}$ and $\mathcal{W}_{5d}^{\text{np}}$ are actually the same function but written in terms of $e^{\hbar}$ and $e^{-1/\hbar}$ respectively. This observation makes manifest the ``$S$-duality'' invariance of \eqref{BAEexact}, i.e. invariance under the exchange $\hbar \leftrightarrow 1/\hbar$. \\
It is therefore clear that a naive extension of Bethe/Gauge correspondence to five dimensions along the lines we described earlier in this Section is incorrect, or at least incomplete. Let us try to understand why, and how the problems we encountered can or cannot be solved. 

\subsection{Bethe/Gauge correspondence on squashed $S^3$ and $S^5$} \label{subs2.4}

The first point we want to stress is that the instanton partition function in five dimensions is only defined for generic, not general, Omega background parameters $\epsilon_1$, $\epsilon_2$: with this we mean that it presents divergences for some specific values of these parameters (which we call ``non-generic''). In particular, in the NS limit the partition function is not well defined precisely in the case of $\epsilon_1$ corresponding to $\hbar$ real; nevertheless at the level of ADHM quiver quantum mechanics this is expected since at these non-generic values there appear additional zero modes. It is probably because of this reason that the Bethe/Gauge correspondence proposal has not been explicitly extended to five dimensions in \cite{2010maph.conf..265N}. This actually seems to be a more general problem regarding Omega background in odd number of dimensions, since also three dimensional vortex partition functions present the same feature. One could therefore imagine the problems arising in \eqref{BAE} are coming from the failure of Omega background in being a good regularization of the partition functions in $\mathbb{R}^4 \times S^1_R$ (or $\mathbb{R}^2 \times S^1_R$)\footnote{Of course, this is or is not a problem depending on what one wants to do. The partition function on $\mathbb{R}^4_{\epsilon_1,\epsilon_2} \times S^1_R$ is perfectly fine as it is if interpreted as an index; on the other hand if one is looking for a way to solve relativistic quantum mechanical systems in terms of gauge theories something else needs to be considered.}. \\
If this is the problem, then a natural solution would be to study the partition function of our theory not on flat space in Omega background but on a compact manifold: by definition this partition function, were one able to properly compute it, will be well regularized. With applications to relativistic quantum mechanics in mind, one then has to look for a compact geometry such that the would-be $\hbar$ is real: the simplest option turns out to be the squashed sphere $S^5$, while $S^4 \times S^1$ would correspond to having $\hbar$ imaginary\footnote{As we will see, $q = e^{2\pi i \omega_1 / \omega_2}$ in $S^5$ plays the role of $q = e^{2\pi i \hbar}$ in quantum mechanics; $\omega_1$, $\omega_2$ are geometric parameters of $S^5$ when $\omega_1, \omega_2 \in \mathbb{R}_+$ which leads to $\hbar$ real. Similarly, reality of the geometric parameters of $S^4 \times S^1$ would correspond to $\hbar$ imaginary.}. This is how the proposal for a non-perturbative completion of refined closed topological strings put forward in \cite{2012arXiv1210.5909L} should be interpreted. 
While this idea is certainly as good as simple, it might not be immediate to implement it rigorously. In order to better understand what the problems could be, let us first consider what happens in three dimensions; ours will be a general discussion, the details will be given in the next Section. \\

Let us work on the squashed three-sphere $S^3_{\omega_1, \omega_2}$ defined as
\begin{equation}
\omega_1^2 \vert z_1 \vert^2 + \omega_2^2 \vert z_2 \vert^2 = 1 \;\;\;,\;\;\; z_{1},z_{2} \in \mathbb{C} \;\;\;,\;\;\; \omega_1, \omega_2 \in \mathbb{R}_+ 
\end{equation}
and take for definiteness a 3d $\mathcal{N} = 2$ gauge theory with gauge group $U(1)$ and two chiral multiplets: this is the example we will discuss in more detail in Section \ref{subs3.1}. As we will see there, the partition function in this case reads
\begin{equation}
\mathcal{Z}_{S^3_{\omega_1, \omega_2}} = \int d\sigma e^{\frac{2\pi i \xi \sigma}{\omega_1 \omega_2}} \mathcal{S}_2(\sigma - a/2 \vert \omega_1, \omega_2) \mathcal{S}_2(\sigma + a/2 \vert \omega_1, \omega_2) 
\label{inaux1}
\end{equation}
The meaning of the various parameters entering this expression will be given in the next Section. 
The function $S_2$ is strictly related to the double sine function (see Appendix \ref{A}), which is defined in terms of a contour integral
. When Im$\,(\omega_1/\omega_2) > 0$ the double sine admits a factorized form in terms of infinite products as in \eqref{product} which could also be used for Im$\,(\omega_1/\omega_2) = 0$ and $\omega_1/\omega_2$ irrational, while for Im$\,(\omega_1/\omega_2) = 0$ and $\omega_1/\omega_2$ rational one has to rely on the contour integral representation from which it is however possible to obtain a finite product representation as done in \cite{Garoufalidis:2014ifa}. We anticipate that $\mathcal{Z}_{S^3_{\omega_1, \omega_2}}$ satisfies the finite-difference equation
\begin{equation}
\left( e^{-i \omega_1 \partial_{\xi}} + e^{i \omega_1 \partial_{\xi}} - e^{-2\pi \xi/\omega_2} \right) \mathcal{Z}_{S^3_{\omega_1, \omega_2}} = \left( \mu^{1/2} + \mu^{-1/2} \right) \mathcal{Z}_{S^3_{\omega_1, \omega_2}} \label{inaux2}
\end{equation}
as well as its modular dual
\begin{equation}
\left( e^{-i \omega_2 \partial_{\xi}} + e^{i \omega_2 \partial_{\xi}} - e^{-2\pi \xi/\omega_1} \right) \mathcal{Z}_{S^3_{\omega_1, \omega_2}} = \left( \widetilde{\mu}^{1/2} + \widetilde{\mu}^{-1/2} \right) \mathcal{Z}_{S^3_{\omega_1, \omega_2}} \label{inaux2dual}
\end{equation}
For $\omega_1, \omega_2 \in \mathbb{R}_+$ (which corresponds to having $\hbar$ real) and all other parameters real the operators on the left hand side of \eqref{inaux2}, \eqref{inaux2dual} are self-adjoint and \eqref{inaux1} is a well-defined eigenfunction of them. This is in fact the solution to the relativistic 2-particle open Toda chain given in \cite{Kharchev:2001rs}. \\
Although we have no reason to do so since everything is consistent and the problem is solved, let us now consider the case in which $\omega_1$, $\omega_2$ are complex, say $\overline{\omega}_1 = \omega_2$; this means that technically speaking we will no longer be on the squashed $S^3$ geometry. Then we can use \eqref{product} and evaluate the integral; as we will see, the result roughly factorizes as
\begin{equation}
\mathcal{Z}_{S^3_{\omega_1, \omega_2}} \;\;\sim \;\; Z_{\mathbb{R}^2_{\omega_1} \times S^1_{\omega_2}} \widetilde{Z}_{\mathbb{R}^2_{\omega_2} \times S^1_{\omega_1}} \label{inaux3}
\end{equation}
in terms of two copies of the partition function on flat space $\mathbb{R}^2_{\epsilon} \times S^1_R$ in the presence of Omega background; the two copies (untilded and tilded) correspond to the $\mathbb{R}^2_{\epsilon} \times S^1_R$ at the North and South poles of $S^3_{\omega_1, \omega_2}$ respectively and are related by the exchange $\omega_1 \leftrightarrow \omega_2$. The function $Z_{\mathbb{R}^2_{\omega_1} \times S^1_{\omega_2}}$ by itself satisfies \eqref{inaux2} but it presents quasi-constant ambiguities, i.e. it is only defined up to functions periodic in $\omega_1$; the easiest solution to this problem involves considering the action of the modular dual operator \eqref{inaux2dual} which will fix in an unique way the quasi-constant to be $\widetilde{Z}_{\mathbb{R}^2_{\omega_2} \times S^1_{\omega_1}}$.
At this point we have that the combination $Z_{\mathbb{R}^2_{\omega_1} \times S^1_{\omega_2}} \widetilde{Z}_{\mathbb{R}^2_{\omega_2} \times S^1_{\omega_1}}$ is an eigenfunction of our operator as well as its modular dual for $\omega_1$, $\omega_2$ complex, but we have to remember that the operators \eqref{inaux2}, \eqref{inaux2dual} are not self-adjoint by themselves unless the $\omega_i$'s are real: only linear combinations of \eqref{inaux2} and \eqref{inaux2dual} are self-adjoint, as discussed at the end of Section \ref{subs2.2}, and the combination \eqref{inaux3} is a good eigenfunction for them. \\
Let us now try to take the $\omega_i \rightarrow \mathbb{R}_+$ limit from our current situation. For $\omega_i$'s real $Z_{\mathbb{R}^2_{\omega_1} \times S^1_{\omega_2}}$ and $\widetilde{Z}_{\mathbb{R}^2_{\omega_2} \times S^1_{\omega_1}}$ are separately ill-defined and have an infinitely dense set of poles in $\omega_1$, $\omega_2$, so we will have to consider them together. Next, by an analytic continuation in $\widetilde{q}$ we can for example bring $\widetilde{Z}_{\mathbb{R}^2_{\omega_2} \times S^1_{\omega_1}}$ from the numerator to the denominator in \eqref{inaux3} (something along the lines of \eqref{continuation}) to arrive at the combination\footnote{Expressions like $Z_{\mathbb{R}^2_{\omega_1} \times S^1_{\omega_2}} \widetilde{Z}_{\mathbb{R}^2_{\omega_2} \times S^1_{\omega_1}}$ are intended as written in terms of $q = e^{2\pi i \omega_1/\omega_2}$ and $\widetilde{q} = e^{2\pi i \omega_2 / \omega_1}$ with Im$\,\omega_1/\omega_2 > 0$, which means $\vert q \vert < 1$ and $\vert \widetilde{q} \vert > 1$. In order to approach Im$\,\omega_1/\omega_2 = 0$ we first have to analytically continue $\widetilde{Z}_{\mathbb{R}^2_{\omega_2} \times S^1_{\omega_1}}$ in $\widetilde{q}$; here and in the following we denote the resulting expression, written in terms of $\vert \widetilde{q}^{-1} \vert < 1$, by $1/\widetilde{Z'}_{\mathbb{R}^2_{\omega_2} \times S^1_{\omega_1}} $. \label{fn}}
\begin{equation}
\mathcal{Z}_{S^3_{\omega_1, \omega_2}} \;\;\sim \;\; Z_{\mathbb{R}^2_{\omega_1} \times S^1_{\omega_2}} / \widetilde{Z'}_{\mathbb{R}^2_{\omega_2} \times S^1_{\omega_1}} \label{inaux4}
\end{equation}
It would actually be quite hard to perform such an analytic continuation with the expressions provided by gauge theory; this is why at this point it is better to rewrite our expressions in an open topological strings like form. In this form it is also easy to show that \eqref{inaux4} is free of poles at $\omega_1/\omega_2$ rational; in order to see that \eqref{inaux4} satisfies \eqref{inaux2}, \eqref{inaux2dual} at $\omega_1/\omega_2$ rational one has or to reconstruct double sine functions from open topological strings or to perform a careful limit along the lines of \cite{Garoufalidis:2014ifa} (see also \cite{Cecotti:2010fi}). \\
The conclusion is the following. The partition function on $S^3_{\omega_1, \omega_2}$ \eqref{inaux1}, which we know how to compute, is a good eigenfunction for the quantum operator \eqref{inaux2} and its modular dual for $\omega_1, \omega_2$ real (or for linear combinations of these operators for $\omega_1, \omega_2$ complex with $\overline{\omega}_1 = \omega_2$) while the partition function on flat space in Omega background $Z_{\mathbb{R}^2_{\omega_1} \times S^1_{\omega_2}}$ can never be a good eigenfunction for \eqref{inaux2} by itself. If we want to reconstruct the true solution \eqref{inaux1} from $Z_{\mathbb{R}^2_{\omega_1} \times S^1_{\omega_2}}$ we have to do a few steps. First, we go to $\omega_1, \omega_2$ complex with $\overline{\omega}_1 = \omega_2$ and invoke modular duality; this brings us to  
\eqref{inaux3}. Second, we perform an analytic continuation in $\widetilde{q}$ in order to get the more proper expression \eqref{inaux4}: this second step involves rewriting our formulae in open topological strings like form, which also allows us to show that \eqref{inaux4} is free of poles at $\omega_1/\omega_2$ rational. Lastly, we take the limit $\omega_1/\omega_2 \rightarrow \mathbb{R}_+$; in order to clearly see that \eqref{inaux4} satisfies a finite difference equation (and its dual) also at $\omega_1/\omega_2$ rational we express everything in terms of double sine functions or we take a careful limit $\omega_1/\omega_2 \rightarrow \mathbb{Q}_+$ to get to expressions like the ones in \cite{Garoufalidis:2014ifa}. \\

\begin{figure}[h]
\centering
\includegraphics[width=0.75\textwidth]{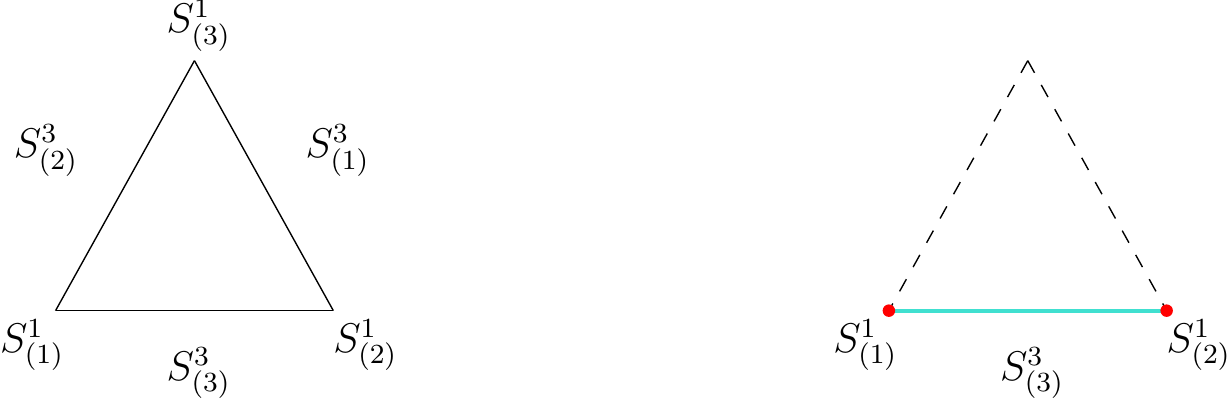}
\caption{Left: $S^5$ as an $(S^1)^3$ fibration over a triangle. Right: the $\omega_3 \rightarrow 0$ limit.} \label{Fig:triangle}
\end{figure}

We can now move to the squashed $S^5$ case. This space has a geometry
\begin{equation}
\omega_1^2 \vert z_1 \vert^2 + \omega_2^2 \vert z_2 \vert^2 + \omega_3^2 \vert z_3 \vert^2 = 1 \;\;\;,\;\;\; z_{1},z_{2},z_{3} \in \mathbb{C} \;\;\;,\;\;\; \omega_1, \omega_2, \omega_3 \in \mathbb{R}_+ 
\end{equation}
and can alternatively thought of as an $(S^1)^3$ fibration over a triangle (see Figure \ref{Fig:triangle}), where the vertices correspond to the fixed circles $S^1_{(i)}$ with respect to the $U(1)^3$ isometries of squashed $S^5$ and the edges correspond to squashed three-spheres. The partition function of 5d $\mathcal{N}=1$ theories on this $S^5$ has been discussed in some detail in \cite{Kim:2012ava,Kim:2012qf,Kallen:2012cs,Hosomichi:2012ek,Kallen:2012va,Imamura:2012xg,Imamura:2012bm}. 
As far as the 1-loop part of this partition function is concerned, the story is very similar to what we already saw for $S^3_{\omega_1, \omega_2}$: the 1-loop part is given by triple sine functions, whose contour integral definition is valid for $\omega_1, \omega_2, \omega_3$ real, while allowing $\omega_1, \omega_2, \omega_3$ complex the triple sine will factorize into three copies of the 1-loop part of the 5d instanton partition function on flat space (one for each fixed $S^1_{(i)}$). However, in this second case we can no longer talk about partition function on $S^5$ but of an analytic continuation of it. For the instanton part (which is not present in 3d) the situation is less clear: 
the expectation is that the integrand of $\mathcal{Z}_{S^5}$ will factorize into three copies of the complete (1-loop + instanton) flat space partition function $Z_{\mathbb{R}^2_{\epsilon_1} \times \mathbb{R}^2_{\epsilon_2} \times S^1_{R}}$, each copy corresponding to one of the vertices of the triangle, with Omega background parameters determined as in Table \ref{tabo}; this would lead to
\begin{equation}
\mathcal{Z}_{S^5} \sim \int [da] \, 
Z_{\mathbb{R}^2_{\omega_1} \times \mathbb{R}^2_{\omega_3} \times S^1_{\omega_2}}^{(1)}
Z_{\mathbb{R}^2_{\omega_3} \times \mathbb{R}^2_{\omega_2} \times S^1_{\omega_1}}^{(2)}
Z_{\mathbb{R}^2_{\omega_2} \times \mathbb{R}^2_{\omega_1} \times S^1_{\omega_3}}^{(3)} \label{whoknows}
\end{equation}
By performing an analytic continuation in the $\omega_i$'s, the integrand can be schematically rewritten in the form\footnote{Again, it is by no means obvious that such an analytic continuation can be performed by looking at the explicit expressions of the 5d partition function; this is something which is better seen by rewriting instanton partition functions, without and with defects, in (closed or open) topological strings-like form. We will see an example of this in Section \ref{subs3.6}.}
\begin{equation}
Z_{\mathbb{R}^2_{\omega_1} \times \mathbb{R}^2_{\omega_3} \times S^1_{\omega_2}}^{(1)} \Big/ \left(
{Z'}_{\mathbb{R}^2_{\omega_3} \times \mathbb{R}^2_{\omega_2} \times S^1_{\omega_1}}^{(2)}
{Z'}_{\mathbb{R}^2_{\omega_2} \times \mathbb{R}^2_{\omega_1} \times S^1_{\omega_3}}^{(3)} \right) \label{ext}
\end{equation}
where the primed functions are intended along the lines of what we said in Footnote \ref{fn}. 
\renewcommand\arraystretch{1.4}
\begin{table}[h]
\begin{center}
\begin{tabular}{|c|c|c|c|}
\hline           & $-i\epsilon_1$ & $-i\epsilon_2$ & $1/R$ \\
\hline $S^1_{(1)}$ &   $\omega_1$   & $\omega_3$     & $\omega_2$ \\ 
       $S^1_{(2)}$ &   $\omega_3$   & $\omega_2$     & $\omega_1$ \\ 
       $S^1_{(3)}$ &   $\omega_2$   & $\omega_1$     & $\omega_3$ \\ 
\hline
\end{tabular} \caption{} \label{tabo}
\end{center}
\end{table}
\renewcommand\arraystretch{1} 
The proposal in \cite{2012arXiv1210.5909L} is that \eqref{ext} is a good non-perturbative completion of $Z_{\mathbb{R}^2_{\omega_1} \times \mathbb{R}^2_{\omega_3} \times S^1_{\omega_2}}^{(1)}$, or equivalently of refined closed topological strings: this is because the integrand of the $S^5$ partition function will always be well-defined since we are on compact space. This is formally correct, but there are a few points one has to take into account.
First of all, no one really properly computed the instanton part of the $S^5$ partition function; we will have to assume \eqref{whoknows} is correct and see what this implies. 
Second, instanton series in $Q_{5d}$ may not be convergent (they are asymptotic for $\omega_i$'s complex and convergent for $\omega_i$'s real if we are not hitting a pole): problems of convergence can be solved by re-expanding the instanton partition function in a closed topological strings form around the large volume point. Third, expressions like the integrand of \eqref{whoknows} are only valid for $\omega_i$'s complex, that is outside from the $S^5$ geometry, and one should show that they are pole-free for $\omega_i$'s real; in order to see this it would again be more convenient to write our expression in topological strings like form. Cancellation of poles has in fact been shown to happen in the particular case $\omega_3 \rightarrow 0$: as we see from Table \ref{tabo} this corresponds to an NS limit for the flat 5d theories living on the first two fixed points, while for the third one the instanton contribution disappears. We therefore roughly remain with two copies of the NS limit of the flat 5d partition function, related by the exchange $\omega_1 \leftrightarrow \omega_2$; it turns out that the total twisted effective superpotential now contains two pieces, and the corresponding supersymmetric vacua equations are free of poles and compatible with the ones in \eqref{BAEexact} as shown in \cite{2015arXiv150704799H}\footnote{Although it does not happen for 5d $SU(2)$, in general cancellation of poles involves an imaginary shift of the Kahler moduli / real scalar fields $a_i$ due to a $B$-field. Such a shift is known to be present in the 5d partition function, although we are not aware of any reference on this technical point.}. The symmetry $\omega_1 \leftrightarrow \omega_2$ is nothing else but the ``$S$-duality'' invariance of \eqref{BAEexact}, which in the squashed $S^5$ setting has a very natural explanation. This provides support to the proposal of considering \eqref{ext} as the non-perturbatively completed partition function; let us however stress that in order to see analytic properties, pole cancellation and convergence it is necessary to rewrite \eqref{ext} in topological strings form, although the computation of \eqref{ext} and related observables may be easier to perform in gauge theory. \\

In the following we will analyse further the proposal of \cite{2012arXiv1210.5909L} by considering codimension two and four defects in the squashed $S^5$ geometry, in a generalization of the setting in Table \ref{tab1}. We will closely follow the set-up of \cite{Bullimore:2014upa}. We will place our codimension two monodromy defect, thought of as the 3d theory of Figure \ref{Fig:figgeneral}, on one of the three squashed 3-spheres inside $S^5$, say $S^3_{(3)}$; the codimension four defects will then be Wilson loops wrapping $S^{1}_{(1)}$ and $S^{1}_{(2)}$. As explained in \cite{Bullimore:2014upa}, the integrand of the $S^5$ partition function in the presence of codimension two defects will conjecturally be of the form
\begin{equation}
\mathcal{Z}_{S^3_{\omega_1, \omega_2} \subset S^5} \;\;\sim\;\; 
\, Z^{(1)}_{3d/5d} Z^{(2)}_{3d/5d} Z^{(3)}_{5d} \label{S5def}
\end{equation}
The integrand contains two copies of the flat 5d partition function with monodromy defect and one copy of the usual 5d partition function. In the $\omega_3 \rightarrow 0$ limit the last one becomes trivial, while the first two reduce to their NS limit and we remain with an integrand like
\begin{equation}
Z^{(1)\,NS}_{3d/5d} Z^{(2)\,NS}_{3d/5d} \label{combo}
\end{equation}
which can be analytically continued to
\begin{equation}
Z^{(1)\,NS}_{3d/5d} \Big/ {Z'}^{(2)\,NS}_{3d/5d} \label{combo2}
\end{equation}
Clearly, $Z^{(1)\,NS}_{3d/5d}$ and $Z^{(2)\,NS}_{3d/5d}$ are related by the exchange $\omega_1 \leftrightarrow \omega_2$ (``$S$-duality''). The modular double structure of the associated quantum relativistic integrable system has a very natural interpretation in this setting: the twisted chiral ring operators of the two copies of the flat theory in the NS limit will give rise to \textit{two} commuting sets of Hamiltonians $\widehat{H}_k$ and $\widehat{\widetilde{H}}_k$ related by $\omega_1 \leftrightarrow \omega_2$; the combination \eqref{combo} will then be a common eigenfunction for these two sets, with eigenvalues corresponding to NS Wilson loops at $S^{1}_{(1)}$ and $S^{1}_{(2)}$ respectively, again mapped between each other by $\omega_1 \leftrightarrow \omega_2$. Alternatively, these finite-difference equations satisfied by \eqref{combo} can be thought of as Ward identities for line operators at $S^{1}_{(1)}$ and $S^{1}_{(2)}$. The usual caveats have to be taken into account. First, \eqref{combo} may not be convergent as a series in $Q_{5d}$; however we will see that it can be written in terms of refined open topological strings which have better convergence properties. Second, \eqref{combo} is only valid for $\omega_i$'s complex, i.e. outside the $S^5$ geometry: going to open topological strings we can see that it is free of poles at $\omega_1/\omega_2$ real ad rational and we can take the limit $\omega_i \rightarrow \mathbb{R}_+$ (with special care in the case $\omega_i \rightarrow \mathbb{Q}_+$). \\

\begin{figure}[h]
\centering
\includegraphics[width=0.7\textwidth]{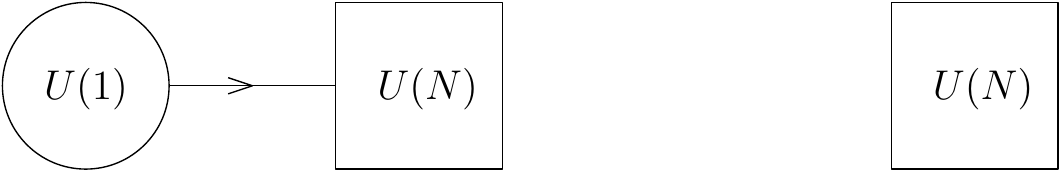}
\caption{Left: simple defect. Right: type IIB $S$-dual to simple defect.} \label{Fig:simplefree}
\end{figure}

We will focus on the 5d $SU(2)$ theory and discuss the eigenfunctions for the associated modular double relativistic 2-particle closed Toda chain. In addition, we will discuss the eigenfunctions of the quantum operator which appears when quantizing the Seiberg-Witten curve: in gauge theory terms this corresponds to consider a different kind of codimension two defect (right side of Figure \ref{Fig:simplefree}), which is simply the $S$-dual (in type IIB strings sense) to a simple monodromy defect. For the special case of $SU(2)$ full and simple defects coincide, and we will study eigenfunctions for the quantized spectral curve and its $S$-dual; as we will see, at the level of integrable systems $S$-duality corresponds to Fourier transformation. In Table \ref{tab11} we give a basic dictionary between the Toda system and the gauge theory on $S^5$ for the convenience of the reader. More details will be given in the next Sections.

\vspace*{1 cm}

\renewcommand\arraystretch{1.5}
\begin{center}
\begin{table}[h]
\begin{tabular}{|c|c|}
\hline \textbf{modular double $N$-particle closed Toda} & \textbf{5d $\mathcal{N}=1$ $SU(N)$ theory on $S^5$ ($\omega_3 \rightarrow 0$)} \\ 
\hline (quantized) spectral curve & (quantized) Seiberg-Witten curve (copy $(1)$) \\ 
\hline (quantized) dual spectral curve & (quantized) Seiberg-Witten curve (copy $(2)$) \\ 
\hline exact Bethe Ansatz Equations & $S^5$ supersymmetric vacua equations \\
\hline quantum Hamiltonians $\widehat{H}_k$ & twisted chiral ring operators (copy $(1)$) \\
\hline dual quantum Hamiltonians $\widehat{\widetilde{H}}_k$ & twisted chiral ring operators (copy $(2)$) \\ 
\hline eigenvalues $E^{(k)}_{\vec{n}}$ & Wilson loops $\langle W^{SU(N)}_{\Lambda^k} \rangle^{NS}\Big\vert_{\vec{n}}$ (copy $(1)$) \\
\hline dual eigenvalues $\widetilde{E}^{(k)}_{\vec{n}}$ & Wilson loops $\langle \widetilde{W}^{SU(N)}_{\Lambda^k} \rangle^{NS}\Big\vert_{\vec{n}}$ (copy $(2)$) \\
\hline $\widehat{H}_k$, $\widehat{\widetilde{H}}_k$ eigenfunctions $\psi_{\vec{n}}(x_1, \ldots, x_N)$ & 
full defect $\langle Z^{(1)}_{3d/5d} Z^{(2)}_{3d/5d}(x_1, \ldots, x_N) \rangle^{NS} \Big\vert_{\vec{n}}$ \\
\hline coordinates $(x_1, \ldots, x_N)$ & monodromy (FI) parameters $(m_1, \ldots, m_N)$ \\
\hline boundary condition parameters $Q$, $\widetilde{Q}$ & instanton counting parameters $Q_{5d}$, $\widetilde{Q}_{5d}$ \\
\hline quantum curves eigenfunctions $\mathcal{Q}_{\vec{n}}(x)$ & 
$S$-dual to simple defect $\langle Z^{(1)}_{3d/5d} Z^{(2)}_{3d/5d}(x) \rangle^{NS} \Big\vert_{\vec{n}}$ \\
\hline Fourier transformed $\mathcal{Q}_{\vec{n}}(x)$  & simple defect partition function \\
\hline
\end{tabular} \caption{} \label{tab11}
\end{table}
\end{center}
\renewcommand\arraystretch{1}

\section{Relativistic 2-particle Toda} \label{sec3}

In this Section we will show how one can construct eigenfunctions and eigenvalues for the modular double relativistic 2-particle Toda (both open and closed) in gauge theory, in a way that takes into account also quantum mechanical instanton contributions (i.e. non-perturbative corrections in $\hbar$).
As mentioned in the previous Section, our discussion will develop further the ideas contained in \cite{2012arXiv1210.5909L} by following the set-up of \cite{Bullimore:2014upa}. We will first collect the gauge theory computations relative to the open and closed Toda chains as well as to their quantized spectral curve, in the meanwhile discussing Fourier transformation of the wave-function; later we will move to open topological strings formalism in order to check regularity of the solution. \\
A complete, well-defined solution to the open 2-particle chain (and more in general to the open $N$-particle chain) has been given in \cite{Kharchev:2001rs} in terms of representation theory of the modular double $U_q(\mathfrak{sl}(2,\mathbb{R})) \otimes U_{\widetilde{q}}(\mathfrak{sl}(2,\mathbb{R}))$ as well as in the context of Quantum Inverse Scattering Method. As we will see, this solution exactly coincides with the gauge theoretical one. 


\subsection{A toy model} \label{subs3.0}

Before moving to more complicated cases, it is actually better to pause a moment and understand the problems we will encounter in a toy model strictly related to the quantum dilogarithm function which, as we will see, is the basic ingredient needed in the rest of the paper. 
Suppose we want to look for eigenfunctions of the finite-difference equation
\begin{equation}
e^{-i \omega_1 \partial_x} + e^{2\pi x/\omega_2} - 1 = 0 \label{qqq}
\end{equation}
It is easy to see that $(q e^{2\pi x/\omega_2};q)_{\infty}$ with $q=e^{2\pi i \omega_1/\omega_2}$ is annihilated by our finite-difference operator:
\begin{equation}
\left(e^{-i \omega_1 \partial_x} + e^{2 \pi x/\omega_2} - 1 \right) (q e^{2\pi x/\omega_2};q)_{\infty} = 0
\end{equation}
However this solution is not satisfactory for many reasons. First of all it presents quasi-constant ambiguities: we can multiply it by any periodic function of $x$ with period $i \omega_1$ and it will still be a solution. This would make the definition of norm of an eigenfunction meaningless, is in contrast with the usual quantum mechanical case in which the only ambiguity is an overall constant. Moreover, the function $(q e^{2\pi x/\omega_2};q)_{\infty}$ is only convergent when $\vert q \vert < 1$; if $\vert q \vert > 1$ we can use the analytic continuation formula \eqref{continuation} and arrive at $(e^{2\pi x/\omega_2};q^{-1})^{-1}_{\infty}$, but we would like an expression which is valid at $\vert q \vert = 1$ (i.e. $\omega_1, \omega_2$ real) since only in this case our finite-difference operator is self-adjoint. Unfortunately our solution is ill-defined at $\vert q \vert = 1$ since it presents poles when $\omega_1/\omega_2$ is rational, as we can see from \eqref{continuation} i.e.
\begin{equation}
(q e^{2\pi x/\omega_2};q)_{\infty} = \text{exp} \left( \sum_{k \geqslant 1} \dfrac{e^{2\pi k x/\omega_2}}{k(1 - q^{-k})} \right) = (e^{2\pi x/\omega_2};q^{-1})^{-1}_{\infty}
\end{equation}
All these problems can be solved if we require the true wave-function to satisfy a second, ``dual'' finite-difference equation
\begin{equation}
e^{-i \omega_2 \partial_x} + e^{2\pi x/\omega_1} - 1 = 0 \label{rrr}
\end{equation}
which is just our previous operator with the exchange $\omega_1 \leftrightarrow \omega_2$. If we define $\widetilde{q}=e^{2\pi i \omega_2/\omega_1}$, surely $(\widetilde{q} e^{2\pi x/\omega_1};\widetilde{q})_{\infty}$ and $(e^{2\pi x/\omega_1};\widetilde{q}^{-1})^{-1}_{\infty}$ will satisfy \eqref{rrr} for $\vert \widetilde{q} \vert < 1$ and $\vert \widetilde{q} \vert > 1$ respectively, while our previous solution will be a quasi-constant for \eqref{rrr}. Therefore we can imagine the combination
\begin{equation}
(q e^{2\pi x/\omega_2};q)_{\infty} (\widetilde{q} e^{2\pi x/\omega_1};\widetilde{q})_{\infty}
\end{equation}
will be a good eigenfunctions for the operators \eqref{qqq}, \eqref{rrr}; nevertheless if we choose $\vert q \vert < 1$ this implies $\vert \widetilde{q} \vert > 1$, so our combination has to be written more properly as
\begin{equation}
\dfrac{(q e^{2\pi x/\omega_2};q)_{\infty}}{(e^{2\pi x/\omega_1};\widetilde{q}^{-1})_{\infty}} \label{lasagne}
\end{equation}
It is easy to show that this combination is regular even at $\vert q \vert = \vert \widetilde{q} \vert = 1$ since the poles at $\omega_1/\omega_2$ rational cancel: in fact this is nothing else but the quantum dilogarithm \eqref{pizza}. Nevertheless, the representation \eqref{lasagne} is properly only valid when Im$\,(\omega_1/\omega_2) > 0$; the limit Im$\,(\omega_1/\omega_2) \rightarrow 0$ can be taken without harm when $\omega_1/\omega_2$ is irrational, but more care is needed when $\omega_1/\omega_2$ is rational: we refer to \cite{Garoufalidis:2014ifa,Cecotti:2010fi} for more details on the proper treatment of the rational case. Just as an example, the proper treatment of the case $\omega_1 \rightarrow 1$, $\omega_2 \rightarrow 1$ leads to an expression like \cite{Garoufalidis:2014ifa}
\begin{equation}
\Phi_{1,1}(x) = \text{exp} \left( \dfrac{i}{2\pi} \left[ \text{Li}_2(e^{2\pi x}) + 2\pi x \ln(1-e^{2\pi x}) \right] \right)
\end{equation}
Alternatively, after having realized that \eqref{lasagne} is the quantum dilogarithm one can just work with its contour integral representation which is valid also at $\omega_1/\omega_2$ rational.

\subsection{Open Toda} \label{subs3.1}

We are now ready to discuss Toda chains. As we discussed in Section \ref{sec2}, the relativistic 2-particle Toda chain is naturally associated to the five dimensional $\mathcal{N}=1$ $SU(2)$ gauge theory: this can be seen for example from the spectral curve of the classical Toda chain, which coincides with the Seiberg-Witten curve of our five dimensional theory. Bethe/Gauge correspondence tells us how this relation gets modified when we consider the quantum Toda chain. 
Here we start from the open chain ($Q=0$): since $Q = Q_{5d} = e^{-8\pi^2 R/g^2_{YM}}$, at the level of gauge theory this corresponds to freezing the five dimensional gauge dynamics (i.e. $g_{YM} = 0$), so that we can just focus on the theory living on the 3d defect. We will first discuss the computation on flat space $\mathbb{R}^2_{\epsilon_1} \times S^1_R$ in order to point out its problems; later we explain how to solve these problems by going to curved space $S^{3}_{\omega_1, \omega_2}$. As we will see, our discussion will not contain anything new conceptually with respect to the toy model we considered in Section \ref{subs3.0}. \\

\noindent \textit{Flat space analysis} \\

\noindent In flat space $\mathbb{R}^2_{\epsilon} \times S^1_R$ we already know how to solve the Toda system in gauge theoretical terms: we just need to follow what we said in Section \ref{subs2.3} (see the dictionary given in Table \ref{tab1}) in the $Q_{5d} = 0$ limit. In particular the eigenfunctions of our system will be given by the NS ramified instanton partition function $Z_{3d/5d}$ for a full monodromy defect, in the limit in which the $SU(2)$ five dimensional bulk theory is decoupled from the defect. 
This is expected to coincide with the 3d block $\mathcal{B}_{3d}$ (the ``K-theoretical'' vortex partition function on $\mathbb{R}^2_{\epsilon} \times S^1_R$, or ``K-theoretical'' Givental's function \cite{2001math......8105G}) for the 3d theory corresponding to our full monodromy defect
, which in this case is a 3d $\mathcal{N}=2$ $U(1)$ theory with two fundamental flavours and flavour symmetry $SU(2)$ (as in Figure \ref{Fig:2toda} after decoupling a flavour $U(1)$). The block $\mathcal{B}_{3d}$ can be extracted from factorization \cite{2012JHEP...04..120P} of the squashed $S^3_{\omega_1, \omega_2}$ partition function \cite{Hama:2011ea} 
\begin{equation}
\mathcal{Z}^{(k)}_{S^3_{\omega_1, \omega_2}} = \int d\sigma e^{\frac{2\pi i \xi \sigma}{\omega_1 \omega_2}} e^{-\frac{i k \pi \sigma^2}{\omega_1 \omega_2}} \mathcal{S}_2(\sigma - a/2 \vert \omega_1, \omega_2) \mathcal{S}_2(\sigma + a/2 \vert \omega_1, \omega_2) \label{int2}
\end{equation}
All the details on the integration contour can be found in \cite{Kharchev:2001rs}. Here $k \in \mathbf{Z}$ is a Chern-Simons level, $\xi$ the Fayet-Iliopoulos parameter, $a/2$ the flavour mass in the Cartan of $SU(2)$, $\omega_1$, $\omega_2$ the squashing parameters of $S^3_{\omega_1, \omega_2}$ parameterized as 
\begin{equation}
\omega_1^2 \vert z_1 \vert^2 + \omega_2^2 \vert z_2 \vert^2 = 1 \;\;\;,\;\;\; z_{1},z_{2} \in \mathbb{C} \;\;\;,\;\;\; \omega_1, \omega_2 \in \mathbb{R}_+ \label{metric}
\end{equation}
and $\mathcal{S}_2(\sigma \vert \omega_1, \omega_2)$ is related to the so-called double sine function $s_{\omega_1, \omega_2}(\sigma)$ as
\begin{equation}
\mathcal{S}_2(\sigma \vert \omega_1, \omega_2) = s_{\omega_1, \omega_2}\left(\sigma + i \frac{\omega_1 + \omega_2}{2} \right)
\end{equation}
See Appendix \ref{A} for more details on the double sine function. Let us define
\begin{equation}
z = e^{-2\pi \xi /\omega_2} \;,\; q=e^{2\pi i \omega_1/\omega_2} \;,\; \mu = e^{2 \pi a / \omega_2} 
\end{equation}
and
\begin{equation}
\widetilde{z} = e^{-2\pi \xi / \omega_1} \;,\; \widetilde{q} = e^{2\pi i \omega_2 / \omega_1} \;,\; \widetilde{\mu} = e^{2 \pi a / \omega_1}
\end{equation}
In order to extract the block from \eqref{int2} we need to perform the integral and factorize the result. This has been done in many places in the literature, see for example \cite{2012JHEP...04..120P}. Following this procedure we arrive at
\begin{equation}
\begin{split}
\mathcal{Z}^{(k)}_{S^3_{\omega_1, \omega_2}} & \;=\; e^{- \frac{i \pi k a^2}{4\omega_1 \omega_2}} \mathcal{S}_2(a \vert \omega_1, \omega_2) Z_{3d,1l}^{(1)}Z_{3d}^{(1)} \widetilde{Z}_{3d,1l}^{(1)} \widetilde{Z}_{3d}^{(1)} \\
&\; +\; e^{-\frac{i \pi k a^2}{4\omega_1 \omega_2}} \mathcal{S}_2(-a \vert \omega_1, \omega_2) Z_{3d,1l}^{(2)}Z_{3d}^{(2)} \widetilde{Z}_{3d,1l}^{(2)} \widetilde{Z}_{3d}^{(2)} \label{fact}
\end{split}
\end{equation}
where
\begin{eqnarray}
Z_{3d,1l}^{(1)} & = & \dfrac{\theta(z \mu^{-1/2};q)}{\theta(z;q)\theta(\mu^{-1/2};q)} \label{1l3d2} \\
Z_{3d}^{(1)} & = & \sum_{m \geqslant 0} \dfrac{\left(-z q^{1/2} \mu^{(1+k)/2} \right)^m q^{(1+k)m^2/2}}{(q;q)_m(q\mu ;q)_m} \label{OT2}
\end{eqnarray}
and
\begin{equation}
\theta(z;q) = (z;q)_{\infty}(q z^{-1};q)_{\infty} = \prod_{k=0}^{\infty} (1-zq^k)(1-z^{-1}q^{1+k})
\end{equation}
The expressions for $Z_{3d,1l}^{(2)}$, $Z_{3d}^{(2)}$ are obtained from \eqref{1l3d2}, \eqref{OT2} by exchanging $\mu \leftrightarrow \mu^{-1}$, while those for tilded quantities coincide with \eqref{1l3d2}, \eqref{OT2} written in terms of tilded variables (i.e. $\omega_1 \leftrightarrow \omega_2$). \\
Modulo prefactors that are not relevant for our discussion, the blocks $\mathcal{B}_{3d}^{(1)}$, $\mathcal{B}_{3d}^{(2)}$ associated to the two vacua of the theory are simply given by $Z_{3d,1l}^{(1)}Z_{3d}^{(1)}$ and $Z_{3d,1l}^{(2)}Z_{3d}^{(2)}$ respectively. These blocks are known to satisfy finite-difference equations, which can be recovered from quantizing the twisted chiral ring relations of the 3d theory \cite{Beem:2012mb,Gaiotto:2013bwa,Bullimore:2014awa}. More in detail, in order to obtain the finite-difference operators we proceed as follows. The factorization of the partition function as in \eqref{fact} suggests that we can see $\mathcal{Z}^{(k)}_{S^3_{\omega_1, \omega_2}}$ as gluing two copies of the partition function on $\mathbb{R}^2_{\epsilon_1} \times S^1_R$ (i.e. two copies of $\mathcal{B}_{3d}$), the first one corresponding to the North pole ($\epsilon_1 \propto \omega_1$, $R \propto 1/\omega_2$) and the second one to the South pole ($\epsilon_1 \propto \omega_2$, $R \propto 1/\omega_1$). We then consider the Coulomb branch of the 3d theory at, say, the North pole; this can be described in terms of the 3d twisted effective superpotential $\mathcal{W}_{\text{eff}}$\footnote{In order to avoid confusion, we stress that this $\mathcal{W}_{\text{eff}}$ is not related to the one in \eqref{YY5d}.}. In our case $\mathcal{W}_{\text{eff}}$ reads
\begin{equation}
\mathcal{W}_{\text{eff}} = \frac{\xi \sigma}{\omega_2} + \omega_2\ell(\sigma - a/2 \vert \omega_2) + \omega_2\ell(\sigma + a/2 \vert \omega_2) - \dfrac{k}{2} \frac{\sigma^2}{\omega_2}
\end{equation}
where $\ell(x \vert \omega_2)$ is a function such that 
\begin{equation}
2 \pi \omega_2 \partial_x \ell(x \vert \omega_2) = \ln\left[2 \sinh \left( \frac{\pi x}{\omega_2} \right) \right]
\end{equation}
The Coulomb branch vacua can be obtained by extremizing $\mathcal{W}_{\text{eff}}$, i.e. they are the solutions to 
\begin{equation}
\text{exp} \left( 2 \pi \dfrac{\partial \mathcal{W}_{\text{eff}}}{\partial \sigma} \right) = 1
\end{equation}
that is
\begin{equation}
e^{2\pi \sigma / \omega_2} + e^{-2\pi \sigma / \omega_2} - e^{2\pi k \sigma /\omega_2} e^{-2\pi \xi / \omega_2} = \mu^{1/2} + \mu^{-1/2} \label{cl2}
\end{equation}
If we now define the momentum conjugate to $\xi$ as
\begin{equation}
p_{\xi} = 2 \pi \dfrac{\partial \mathcal{W}_{\text{eff}}}{\partial \xi} = 2\pi \sigma / \omega_2
\end{equation}
equation \eqref{cl2} reduces to
\begin{equation}
e^{p_{\xi}} + e^{-p_{\xi}} - e^{k p_{\xi}}e^{-2\pi \xi / \omega_2} = \mu^{1/2} + \mu^{-1/2} \label{cl2p}
\end{equation}
The left hand side of \eqref{cl2p} is just a different, $k$-dependent reparameterization of the classical version 
\begin{equation}
H_1 = \left( 1 + e^{-\frac{2\pi}{\omega_2} x} \right) e^{-\omega_1 p_x} + e^{\omega_1 p_x} \;\;\;,\;\;\; x = x_2 - x_1
\end{equation} 
of the relativistic 2-particle open Toda Hamiltonian \eqref{clT} when the center of mass is decoupled, after properly rescaling the parameters; our original parameterization is recovered for $k=-1$. The right hand side of \eqref{cl2p} represents the energy $\mu^{1/2} + \mu^{-1/2}$ of the classical system, which is just the VEV of the $SU(2)$ flavour Wilson loop in the fundamental representation $\langle W^{SU(2)}_{\square} \rangle$ wrapping $S^1_R$. If we now quantize $[\xi, p_{\xi}] = i \omega_1$, equation \eqref{cl2p} reproduces the quantum Toda Hamiltonian $\widehat{H}_1$ \eqref{clT} for $k=-1$, or other parameterizations of it for different values of $k$:
\begin{equation}
\widehat{H}_1^{(k)} = e^{-i \omega_1 \partial_{\xi}} + e^{i \omega_1 \partial_{\xi}} - q^{k/2} e^{-2\pi \xi /\omega_2} e^{-i \omega_1 k \partial_{\xi}} \label{Toda2mod}
\end{equation}
The associated quantum problem will therefore be finding the $k$-dependent (and $a$-dependent) function $\psi^{(k)}_{a}(\xi)$ such that
\begin{equation}
\widehat{H}_1^{(k)} \psi^{(k)}_a(\xi) = (\mu^{1/2} + \mu^{-1/2})\psi_a^{(k)}(\xi)
\end{equation}
Other equivalent formulations are
\begin{equation}
\psi_a^{(k)}(\xi - i \omega_1) + \psi_a^{(k)}(\xi + i \omega_1) - q^{k/2} e^{-2\pi \xi /\omega_2} \psi_a^{(k)}(\xi - i k \omega_1) = 
(\mu^{1/2} + \mu^{-1/2})\psi_a^{(k)}(\xi) \label{eqfin}
\end{equation}
and
\begin{equation}
\left( p_z + p_z^{-1} - z \, q^{k/2} p_z^{k} \right) \psi_a^{(k)}(\xi) = (\mu^{1/2} + \mu^{-1/2}) \psi_a^{(k)}(\xi) 
\end{equation}
where $p_z$ acts as $p_z z = q z$. One can easily check that the combinations $Z_{3d,1l}^{(i)}Z_{3d}^{(i)}$, $i = 1,2$ in \eqref{fact}, which should correspond to the partition function of our Figure \ref{Fig:2toda} theory on flat space $\mathbb{R}^2_{\epsilon_1} \times S^1_R$, satisfy \eqref{eqfin} at least formally. As pointed out in \cite{Bullimore:2014awa}, this was also known in the mathematical literature in the context of quantum K-theory on flag manifolds \cite{2001math......8105G}; here we basically recovered the same result in gauge theory language. \\

Very often in the gauge theory literature $Z_{3d,1l}^{(i)}Z_{3d}^{(i)}$ are considered as the eigenfunctions of the quantum relativistic 2-particle open Toda chain, or at least as formal eigenfunctions. Unfortunately, this does not make much sense in relativistic quantum mechanics, for the same reasons we discussed in Section \ref{subs3.0}:
\begin{itemize}
\item First, $\omega_1 / \omega_2$ is related to $\hbar$ and therefore all values of $\omega_1 / \omega_2 \in \mathbb{R}_+$ should be allowed; actually these should be the only values one must consider, since the squashed $S^3_{\omega_1, \omega_2}$ metric \eqref{metric} is only defined in this range and the quantum operator \eqref{Toda2mod} is self-adjoint only for $\hbar$ real. Nevertheless, it is easy to see that $Z_{3d}^{(i)}$ in \eqref{OT2} is ill-defined exactly for $\vert q \vert = 1$, in particular it has poles when $\omega_1 / \omega_2 = r/s$ with $r,s \in \mathbb{N}$.
\item Second, eigenfunctions in quantum mechanics are defined modulo at most an overall constant, which can later be fixed to the appropriate value. On the other hand in our case, since we are dealing with a relativistic system which involves finite-difference operators $e^{-i \omega_1 \partial_{\xi}}$, $Z_{3d}^{(i)}$ is only defined up to \textit{quasi-constants}, i.e. functions periodic in $i \omega_1$. For example, the function $\widetilde{Z}_{3d}^{(i)}$ is a quasi-constant with respect to the finite-difference operator $e^{-i \omega_1 \partial_{\xi}}$.
\end{itemize} 
This is telling us that if we want to give our 3d defect theory an interpretation in the context of relativistic quantum integrable systems, it is not enough to consider it on flat space $\mathbb{R}^2_{\epsilon_1} \times S^1_R$. Nevertheless, everything becomes consistent if we move to curved space, in particular to the squashed three-sphere $S^3_{\omega_1, \omega_2}$. \\

\noindent \textit{Curved space analysis} \\

\noindent As we mentioned in Section \ref{subs2.4}, the solution to the problems we just pointed out arises from properly taking into account the existence of the \textit{dual} Toda system as done in \cite{Kharchev:2001rs}. The dual system naturally appears if we consider the 3d theory on the whole $S^3_{\omega_1, \omega_2}$ instead of just $\mathbb{R}^2_{\epsilon_1} \times S^1_R$ at the North pole, as noticed for example in \cite{Beem:2012mb,Dimofte:2014zga}: in fact following the same steps as before the $\mathbb{R}^2_{\epsilon_1} \times S^1_R$ at the South pole will give a dual twisted effective superpotential 
\begin{equation}
\widetilde{\mathcal{W}}_{\text{eff}} = \dfrac{\xi \sigma}{\omega_1} + \omega_1 \ell(\sigma - a/2\vert \omega_1) + \omega_1 \ell(\sigma + a/2 \vert \omega_1) - \dfrac{k}{2} \dfrac{\sigma^2}{\omega_1}
\end{equation}
leading to the ($k$-dependent) dual Toda Hamiltonian
\begin{equation}
\widehat{\widetilde{H}}_1^{(k)} = e^{-i \omega_2 \partial_{\xi}} + e^{i \omega_2 \partial_{\xi}} - \widetilde{q}^{k/2} e^{-2 \pi \xi / \omega_1} e^{-i \omega_2 k \partial_{\xi}}
\end{equation}
and the dual quantum problem
\begin{equation}
\widehat{\widetilde{H}}_1^{(k)} \psi_a^{(k)}(\xi) = (\widetilde{\mu}^{1/2} + \widetilde{\mu}^{-1/2}) \psi_a^{(k)}(\xi)
\end{equation}
Clearly $\widetilde{Z}_{3d,1l}^{(i)} \widetilde{Z}_{3d}^{(i)}$ are formal eigenfunctions of the dual quantum problem, but they have the same problems as $Z_{3d,1l}^{(i)} Z_{3d}^{(i)}$. 
The idea is then to look for simultaneous eigenfunctions of $\widehat{H}_1^{(k)}$ and $\widehat{\widetilde{H}}_1^{(k)}$
with eigenvalues $\mu^{1/2} + \mu^{-1/2}$ and $\widetilde{\mu}^{1/2} + \widetilde{\mu}^{-1/2}$ respectively; this condition will fix the quasi-constant ambiguity completely. By noticing that 
\begin{itemize}
\item $\widetilde{Z}_{3d,1l}^{(i)} \widetilde{Z}_{3d}^{(i)}$ is a quasi-constant for $\widehat{H}_1^{(k)}$ since $p_{z} \widetilde{z} = \widetilde{z}$
\item $Z_{3d,1l}^{(i)} Z_{3d}^{(i)}$ is a quasi-constant for $\widehat{\widetilde{H}}_1^{(k)}$ since $\widetilde{p}_{z} z = z$
\end{itemize}
one could think that the combination $Z_{3d,1l}^{(i)} Z_{3d}^{(i)}\widetilde{Z}_{3d,1l}^{(i)} \widetilde{Z}_{3d}^{(i)}$ is a good candidate: in fact by considering the logarithm of this combination one can easily perform an analytic continuation in $\widetilde{q}$ of $\widetilde{Z}_{3d}^{(i)}$ and show that the limit $\omega_1 / \omega_2 \rightarrow \mathbb{R}_+$ is regular, that is poles at $\omega_1 / \omega_2 = r/s$ cancel. 
Otherwise one could just remember that our combination is the same as $\mathcal{Z}^{(k)}_{S^3_{\omega_1, \omega_2}}$ in \eqref{int2}: this is written in terms of double sine functions and is therefore well defined at any $\omega_1, \omega_2 \in \mathbb{R}_+$. We conclude that the true eigenfunction of the open Toda chain is
\begin{equation}
\psi_a^{(k)}(\xi) = \mathcal{Z}^{(k)}_{S^3_{\omega_1, \omega_2}}(\xi; a)
\end{equation}
In fact it is easy to show that, even without performing the integration, $\mathcal{Z}^{(k)}_{S^3_{\omega_1, \omega_2}}$ satisfies
\begin{equation}
\begin{split}
& \left( p_z + p_z^{-1} - z \, q^{k/2} p_z^{k/2} \right) \mathcal{Z}^{(k)}_{S^3_{\omega_1, \omega_2}}(\xi; a) = (\mu^{1/2} + \mu^{-1/2}) \mathcal{Z}^{(k)}_{S^3_{\omega_1, \omega_2}}(\xi; a) \\
& \left( \widetilde{p}_z + \widetilde{p}_z^{-1} - \widetilde{z} \, \widetilde{q}^{k/2} \widetilde{p}_z^{k/2} \right) \mathcal{Z}^{(k)}_{S^3_{\omega_1, \omega_2}}(\xi; a) = (\widetilde{\mu}^{1/2} + \widetilde{\mu}^{-1/2}) \mathcal{Z}^{(k)}_{S^3_{\omega_1, \omega_2}}(\xi; a) \label{eqf}
\end{split}
\end{equation}
This is consistent with what is known by the integrable system community, since equation \eqref{int2} is the same expression for the eigenfunction given in \cite{Kharchev:2001rs} in the so-called $q$-deformed Mellin-Barnes representation. In gauge theory language, equations \eqref{eqf} can be thought of as Ward identities for flavour line operators inserted at North and South poles as discussed in \cite{Dimofte:2011ju,Dimofte:2011py,Dimofte:2014zga} (following \cite{Drukker:2009id,Alday:2009fs,Gaiotto:2010be}). \\

We conclude this section with the following comment. As noticed in \cite{Kharchev:2001rs}, the Toda operators $\widehat{H}_1$, $\widehat{\widetilde{H}}_1$ are self-adjoint when $\omega_1$, $\omega_2$ are real, but if we consider the whole modular double structure of the Toda system more possibilities are allowed: in particular when $\overline{\omega}_1 = \omega_2$ the operators $\widehat{H}_1 + \widehat{\widetilde{H}}_1$ and $i(\widehat{H}_1 - \widehat{\widetilde{H}}_1)$ become self-adjoint and we can study this quantum mechanical system for $\hbar$ complex. Nevertheless, for $\omega_1$ and $\omega_2$ complex our geometry no longer is a squashed three-sphere and we can no longer talk about codimension two defects on $S^3_{\omega_1 \omega_2}$; this analytic continuation in gauge theory is probably related to changing the geometry to $S^2 \times S^1$ or Lens spaces $L(r,1)$. In fact the modular double structure is a common feature of all of these 3d compact spaces, as recently discussed in \cite{Nedelin:2016gwu}. A similar problem of analytic continuation in $\omega_1, \omega_2$, as well as the identification of two possible domains for self-adjoint operators, also appeared in \cite{Dimofte:2014zga} in the context of 3d-3d correspondence and complex Chern-Simons theory which are more closely related to the discussion in \cite{Nedelin:2016gwu}. 

\subsection{Open bispectral dual Toda (quantized spectral curve)} \label{subs3.2}

From now on, let us focus on \eqref{int2} in the case of Chern-Simons level $k=0$: we then have
\begin{equation}
\mathcal{Z}_{S^3_{\omega_1, \omega_2}}(\xi; a) = \int d\sigma e^{\frac{2\pi i \xi \sigma}{\omega_1 \omega_2}}  \mathcal{S}_2(\sigma - a/2 \vert \omega_1, \omega_2) \mathcal{S}_2(\sigma + a/2 \vert \omega_1, \omega_2) \label{int0}
\end{equation}
which as we have just seen is a simultaneous eigenfunction of the two operators
\begin{equation}
\begin{split}
& \left( p_z + p_z^{-1} - z \right) \mathcal{Z}_{S^3_{\omega_1, \omega_2}}(\xi; a) = (\mu^{1/2} + \mu^{-1/2}) \mathcal{Z}_{S^3_{\omega_1, \omega_2}}(\xi; a) \\
& \left( \widetilde{p}_z + \widetilde{p}_z^{-1} - \widetilde{z} \right) \mathcal{Z}_{S^3_{\omega_1, \omega_2}}(\xi; a) = (\widetilde{\mu}^{1/2} + \widetilde{\mu}^{-1/2}) \mathcal{Z}_{S^3_{\omega_1, \omega_2}}(\xi; a)  \label{operAA0}
\end{split}
\end{equation}
with $z = e^{-2\pi \xi/\omega_2}$, $p_z = e^{- i \omega_1 \partial_{\xi}}$. If we compare these operators with the ones coming from the quantization of the spectral curve \eqref{spe2} and its modular dual in the limit $Q \rightarrow 0$, that is
\begin{equation}
\begin{split}
& - e^{- i \omega_1 \partial_x} \mathcal{Q}_a(x) = \left( e^{2\pi x/\omega_2} - E_1 + e^{-2\pi x/\omega_2} \right) \mathcal{Q}_a(x) \\
& - e^{- i \omega_2 \partial_x} \mathcal{Q}_a(x) = \left( e^{2\pi x/\omega_1} - \widetilde{E}_1 + e^{-2\pi x/\omega_1} \right) \mathcal{Q}_a(x)
\end{split}
\end{equation}
or equivalently
\begin{equation}
\begin{split}
& \left( e^{- i \omega_1 \partial_x} + e^{2\pi x/\omega_2} + e^{-2\pi x/\omega_2} \right) \mathcal{Q}_a(x) = E_1 \mathcal{Q}_a(x) \\
& \left( e^{- i \omega_2 \partial_x} + e^{2\pi x/\omega_1} + e^{-2\pi x/\omega_1}  \right) \mathcal{Q}_a(x) = \widetilde{E}_1 \mathcal{Q}_a(x) \label{operAA1}
\end{split}
\end{equation}
we notice that, modulo phases, \eqref{operAA0} and \eqref{operAA1} are (classically) related by the exchange of coordinate and momenta / Fourier transform, that is by a canonical transformation of variables. Equivalently, they can be seen as different quantizations of the same spectral curves according to what we define being coordinate and momenta at the classical level. From the point of view of integrable systems, this is known as \textit{bispectral duality}; the analogue in gauge theory is $S$ duality in type IIB string theory if we see the 3d theory as a defect for a 5d theory engineered via a $(p,q)$-web brane \cite{Gaiotto:2014ina}. \\
It is immediate to write down the simultaneous eigenfunction for \eqref{operAA1}. In fact, $\mathcal{Z}_{S^3_{\omega_1, \omega_2}}(\xi; a)$ is nothing but the Fourier transform of its integrand\footnote{Fourier transformation of relativistic open Toda wave-functions was also discussed in \cite{1064-5632-79-2-388}.}, which will therefore be the eigenfunction we are looking for,
\begin{equation}
\mathcal{Q}_a(x) = \mathcal{S}_2(x - a/2 \vert \omega_1, \omega_2) \mathcal{S}_2(x + a/2 \vert \omega_1, \omega_2) \label{eigopfree}
\end{equation}
with the same eigenvalues
\begin{equation}
E_1 = \mu^{1/2} + \mu^{-1/2} \;\;\;,\;\;\; \widetilde{E}_1 = \widetilde{\mu}^{1/2} + \widetilde{\mu}^{-1/2}
\end{equation}
This is a simple consequence of the property \eqref{property} of the double sine function. Notice that the function \eqref{eigopfree} is nothing else but the partition function on $S^3_{\omega_1, \omega_2}$ of two free chiral multiplets, which is the defect theory of Figure \ref{Fig:simplefree} right ($S$-dual to a simple defect, although when $N=2$ simple and full defects coincide). 
Clearly one could also consider taking two antichiral multiplets instead of chiral ones, both in this Section and in the previous one: we will not consider this explicitly here since everything goes the same way as for the chiral multiplets once the parameters are appropriately identified.

\subsection{Closed Toda} \label{subs3.3}

Having reinterpreted the known solution \cite{Kharchev:2001rs} for the eigenfunctions of the open Toda chain in terms of gauge theory quantities, we can now move on and discuss the closed Toda chain. As we already mentioned, the eigenfunctions for the closed chain are expected to be related to the $SU(2)$ ramified instanton partition function $Z_{3d/5d}$ for a full monodromy defect (Figure \ref{Fig:2toda}) in the NS limit. In the previous subsections we did not need to compute this quantity directly, since when the five dimensional gauge theory is frozen we can just consider the theory living on the 3d defect; this time instead we really have to perform this computation. 
As we said in Section \ref{subs2.4}, gauge theories computations on flat space will only be valid outside the unit circle $\vert q \vert = 1$ and will be given by a perturbative series in the instanton counting parameter $Q_{5d\,}$ which may not be convergent; in this and the next subsection we will limit ourselves to collect the gauge theory results on flat space and $S^5$, leaving the discussion on convergence and pole cancellation for Section \ref{subs3.6}. For the moment we just remark that the NS limit of the VEV of the 5d gauge Wilson loop $\langle W^{SU(2)}_{\square} \rangle$\footnote{Differently from the $\mathcal{N}=1^*$ case considered in \cite{Bullimore:2014awa}, when $\mathcal{N}=1$ the Wilson loop for a $U(1)$ theory is just 1.} is actually a well-defined quantity even along $\vert q \vert = 1$.
All the relevant gauge theory quantities have already been analysed in \cite{Gaiotto:2014ina,Bullimore:2014awa} for theories on $\mathbb{R}^2_{\epsilon_1} \times \mathbb{R}^2 \times S^1_R$; here we will review their results and use them to study the squashed $S^5$ case. \\

Let us start by reviewing the known results on $\mathbb{R}^2_{\epsilon_1} \times \mathbb{R}^2_{\epsilon_2} \times S^1_R$ in the NS limit. The partition function, as well as the vacuum expectation value of a supersymmetric Wilson loop in a fixed representation of the gauge group wrapping $S^1_R$, can be obtained in terms of equivariant characters for the action of global symmetries on the vector spaces entering the ADHM construction of the instanton moduli space. More details on the computation can be found for example in \cite{Gaiotto:2014ina,Bullimore:2014upa,Bullimore:2014awa}; here we will just collect the final result. In particular for $SU(2)$ the NS limit of the Wilson loop in the fundamental representation reads
\begin{equation}
\begin{split}
\langle W_{\square}^{SU(2)} \rangle^{NS} &= \mu^{1/2} + \mu^{-1/2} - Q_{5d\,} (\mu^{1/2} + \mu^{-1/2}) \dfrac{q}{(1 - q \mu)(1 - q \mu^{-1})} \\
& + Q^2_{5d\,} (\mu^{1/2} + \mu^{-1/2}) \dfrac{\left[ (\mu + \mu^{-1}) (1+q+q^2+q^3+q^4) - (3q+4q^2+3q^3) \right]}{(1 - q \mu)^3(1 - q^2 \mu)(1 - q \mu^{-1})^3(1 - q^2 \mu^{-1})} + o(Q^3_{5d\,}) \label{W2f}
\end{split}
\end{equation}
where this time $q = e^{2\pi R \epsilon_1}$, $\mu = e^{2\pi R a}$, $Q_{5d} = e^{-8\pi^2 R/g^2_{5d}}$. An alternative way to obtain \eqref{W2f} could be to consider the NS limit of a particular line defect introduced in \cite{Gomis:2006sb,Tong:2014cha}, which is also known as $qq$-character $\mathcal{X}(w)$ in \cite{Nekrasov:2012xe,Nekrasov:2013xda,Nekrasov:2015wsu,Bourgine:2015szm,Kimura:2015rgi}: in fact this is defined as the generating function of Wilson loops in the $\Lambda^k$ antisymmetric representations in the case of an $SU(N)$ theory
\begin{equation}
\mathcal{X}_N(w) = \sum_{k=0}^{N} (-1)^k w^{\frac{N}{2}-k} \langle W_{\Lambda^k}^{SU(N)} \rangle
\end{equation}
Since the NS limit of the Wilson loops in the antisymmetric representations coincide with the eigenvalues of the relativistic Toda chain, in the same limit $\mathcal{X}_N(w)^{NS}$ is nothing else but our generating function for the eigenvalues $t_N(w)$ \eqref{gfe}.
The techniques to evaluate the $qq$-character via 1d gauged quantum mechanics can be found in \cite{Kim:2016qqs}. In our $SU(2)$ case this simply gives (in the NS limit)
\begin{equation}
\mathcal{X}_2(w)^{NS} = w - \langle W_{\square}^{SU(2)} \rangle^{NS} + w^{-1} \label{qq2}
\end{equation}
with $\langle W_{\square}^{SU(2)} \rangle^{NS}$ as in \eqref{W2f}. \\
The computation of $Z_{5d,(\rho)}$, i.e. the $SU(N)$ partition function in presence of a monodromy defect, is slightly more involved and requires an orbifolding modification of the computation for $Z_{5d}$ as described in \cite{Alday:2010vg,Kanno:2011fw}; technical details for the case at hand can again be found in \cite{Gaiotto:2014ina,Bullimore:2014upa,Bullimore:2014awa}. The final result will be a series expansion in powers of the parameters
\begin{equation}
z_j = e^{-\frac{8\pi^2 R}{g^2_{5d}}(m_{j+1}-m_j)} \;\;\text{for}\;\; j=1,\ldots, r-1 \;\;\;,\;\;\; z_r = Q_{5d\,} e^{-\frac{8\pi^2 R}{g^2_{5d}}(m_{1}-m_r)} \label{mon}
\end{equation}
where the monodromy parameters $(m_1, \ldots, m_r)$ have been introduced in \eqref{mono}. As summarized in Table \ref{tab1}, we are interested in full defects for which $n_j=1$, $j=1, \ldots, N$ $\left(\rho = (1^N)\right)$; in this case (in the NS limit)
\begin{equation}
Z_{5d,(1^N)}^{NS} = \sum_{\sigma = 1}^{N!} Z_{3d,1l}^{(\sigma)} Z_{3d/5d}^{(\sigma)} 
\end{equation}
When the gauge group is $SU(2)$ this reduces to
\begin{equation}
Z_{5d,(1^2)}^{NS} = Z_{3d,1l}^{(1)} Z_{3d/5d}^{(1)} + Z_{3d,1l}^{(2)} Z_{3d/5d}^{(2)} 
\end{equation}
where $Z_{3d,1l}^{(1)}$ basically coincides with \eqref{1l3d2} modulo prefactors\footnote{The partition function $Z_{5d,(1^N),\sigma}^{NS}$ in principle also contains a five dimensional perturbative contribution, but this simplifies when taking the VEV (i.e. when dividing by the five dimensional partition function).} and
\begin{equation}
\begin{split}
Z_{3d/5d}^{(1)} & = 1 - z \dfrac{q \sqrt{\mu}}{(1-q)(1 - q \mu)}
+ z^2 \dfrac{q^3 \mu}{(1-q)(1-q^2)(1 - q \mu)(1 - q^2 \mu)} + o(z^3) \\
& - Q_{5d\,} z^{-1} \dfrac{q \sqrt{\mu^{-1}}}{(1-q)(1 - q\mu^{-1})} 
+ Q_{5d\,} \dfrac{\left[ (q + 2q^2 - q^3) + \mu^{-1} (q^2 - 2 q^3 - q^4) \right]}{2(1-q)^2(1 - q \mu)(1 - q \mu^{-1})^2} + o(Q_{5d\,} z) \\
& + Q^2_{5d\,} z^{-2} \dfrac{q^3 \mu^{-1}}{(1-q)(1-q^2)(1 - q \mu^{-1})(1 - q^2 \mu^{-1})} + o(Q^2_{5d\,} z^{-1}) + o(Q^3_{5d\,}z^{-3}) \label{5ddef2}
\end{split}
\end{equation}
Here $z=z_1$ in \eqref{mon}. The second defect partition function $Z_{3d,1l}^{(2)} Z_{3d/5d}^{(2)}$ can obtained from the above one by exchanging $\mu \leftrightarrow \mu^{-1}$. As we can easily see, \eqref{5ddef2} reduces to the $k=0$ parameterization of \eqref{OT2} in the $Q_{5d} \rightarrow 0$ limit as expected. \\

As noticed in \cite{Gaiotto:2014ina,Bullimore:2014awa}, order by order in $Q_{5d}$ the $SU(2)$ defect partition function $Z_{3d,1l}^{(i)} Z_{3d/5d}^{(i)}$ is an eigenfunction for the relativistic 2-particle closed Toda Hamiltonian in the parameterization
\begin{equation}
\widehat{H}_1 = e^{- \epsilon_1 \partial_{\xi}} + e^{\epsilon_1 \partial_{\xi}} - e^{-2\pi R \xi} - Q_{5d\,} e^{2\pi R \xi}
\end{equation}
with eigenvalue given by $\langle W_{\square}^{SU(2)} \rangle^{NS}$. More explicitly, for 
\begin{equation}
\xi = 4 \pi (m_2 - m_1)/g^2_{5d}
\end{equation}
we have  
\begin{equation}
\left(e^{- \epsilon_1 \partial_{\xi}} + e^{\epsilon_1 \partial_{\xi}} - e^{-2\pi R \xi} - Q_{5d\,} e^{2\pi R \xi} \right) Z_{3d,1l}^{(i)} Z_{3d/5d}^{(i)} = \langle W_{\square}^{SU(2)} \rangle^{NS} Z_{3d,1l}^{(i)} Z_{3d/5d}^{(i)} \label{343}
\end{equation}
or equivalently
\begin{equation}
\left(p_z + p_z^{-1} - z - Q_{5d\,}z^{-1}\right) Z_{3d,1l}^{(i)} Z_{3d/5d}^{(i)} = \langle W_{\square}^{SU(2)} \rangle^{NS} Z_{3d,1l}^{(i)} Z_{3d/5d}^{(i)}
\end{equation}
where $z=e^{-2 \pi R \xi}$, $p_z = e^{- \epsilon_1 \partial_{\xi}}$ and $p_z$ acts as $p_z z = q z$. 
The parameters are identified as in Table \ref{tab2}, where we can see that the 5d instanton counting parameter $Q_{5d\,}$ corresponds to the auxiliary parameter $Q$ for the closed relativistic Toda chain introduced in Section \ref{subs2.1}. By redefining $z \rightarrow -z$ we arrive at the finite-difference operator associated to local $\mathbb{F}_0$ considered in \cite{Grassi:2014zfa}; this operator has been shown in \cite{Kashaev:2015kha,Laptev:2015loa} to be self-adjoint on $L^2(\mathbb{R})$ and with a discrete spectrum, and with inverse of trace class.
\renewcommand\arraystretch{1.4}
\begin{table}
\begin{center}
\begin{tabular}{|c|c|}
\hline \textbf{relativistic 2-particle closed Toda} & \textbf{5d $\mathcal{N} = 1$ $SU(2)$ theory on $\mathbb{R}^2_{\epsilon_1}\times \mathbb{R}^2 \times S^1_R$} \\ 
\hline $q = e^{2\pi i \omega_1/\omega_2} $ & $q = e^{2\pi R \epsilon_1}$ \\ 
\hline $Q = e^{-8\pi^2/(g^2 \omega_2)}$ & $Q_{5d} = e^{-8\pi^2 R/g^2_{5d}}$ \\ 
\hline $e^{-2\pi x/\omega_2}$ & $e^{-2\pi R \xi}$ \\
\hline $x = x_2 - x_1$ & $\xi = 4 \pi (m_2 - m_1)/g^2_{5d}$ \\ 
\hline $\mu = e^{2\pi a / \omega_2}$ & $\mu = e^{2\pi R a}$ \\
\hline $\omega_2$ & $1/R$ \\
\hline $\omega_1$ & $-i \epsilon_1$ \\
\hline
\end{tabular} 
\caption{} \label{tab2}
\end{center}
\end{table}
\renewcommand\arraystretch{1} 

Nevertheless we know that the defect partition function on $\mathbb{R}^2_{\epsilon_1} \times \mathbb{R}^2 \times S^1_R$ cannot be the whole story, since similarly to what we saw in Section \ref{subs3.1} also $Z_{3d/5d}^{(i)}$ has quasi-constant ambiguities, is ill-defined on the unit circle $\vert q \vert = 1$ and has poles for $-i R \epsilon_1 = r/s$ with $r,s \in \mathbb{N}$. We already discussed many times how to fix these problems: we have to require the true eigenfunction to simultaneously be an eigenfunction of the dual closed Toda Hamiltonian with dual eigenvalue $\langle \widetilde{W}_{\square}^{SU(2)} \rangle^{NS}$, i.e. \eqref{W2f} written in terms of tilded variables. Tilded variables are obtained by exchanging $\omega_1 \leftrightarrow \omega_2$ at the level of Toda system; looking at Table \ref{tab2}, this corresponds to map $R \leftrightarrow \frac{1}{-i \epsilon_1}$ on the gauge theory side. In gauge theoretical terms, considering the full modular double Toda system corresponds to place our theory on the $\omega_3 \rightarrow 0$ limit of the squashed $S^5$ as discussed in Section \ref{subs2.4}; this will produce the required completion of the monodromy defect partition function on flat space. Therefore we conclude that, similarly to \eqref{fact} and along the lines of the proposal reviewed in Section \ref{subs2.4}, the function
\begin{equation}
\begin{split}
\mathcal{Z}_{S^3_{\omega_1, \omega_2} \subset S^5} & \;=\; c_1 \, Z_{3d,1l}^{(1)} Z_{3d/5d}^{(1)} \widetilde{Z}_{3d,1l}^{(1)}  \widetilde{Z}_{3d/5d}^{(1)} \\
& \;+\; c_2 \, Z_{3d,1l}^{(2)} Z_{3d/5d}^{(2)} \widetilde{Z}_{3d,1l}^{(2)} \widetilde{Z}_{3d/5d}^{(2)} \label{eig2clo}
\end{split}
\end{equation}
satisfying
\begin{equation}
\begin{split}
& \left( p_z + p_z^{-1} - z - Q_{5d\,}z^{-1} \right) \mathcal{Z}_{S^3_{\omega_1, \omega_2} \subset S^5} = \langle W_{\square}^{SU(2)} \rangle^{NS} \mathcal{Z}_{S^3_{\omega_1, \omega_2} \subset S^5} \\
& \left( \widetilde{p}_z + \widetilde{p}_z^{-1} - \widetilde{z} - \widetilde{Q}_{5d\,}\widetilde{z}^{-1} \right) \mathcal{Z}_{S^3_{\omega_1, \omega_2} \subset S^5} = \langle \widetilde{W}_{\square}^{SU(2)} \rangle^{NS} \mathcal{Z}_{S^3_{\omega_1, \omega_2} \subset S^5} \label{eqfclosed}
\end{split}
\end{equation}
is the proper eigenfunction of the modular double Toda system. Along the unit circle $\vert q \vert = \vert \widetilde{q} \vert = 1$ the story goes as we said in Section \ref{subs3.1}. As discussed in \cite{Kashani-Poor:2016edc}, convergence properties and cancellation of poles in the limit Im$(\omega_1/\omega_2) \rightarrow 0$ are most easily seen if we rewrite \eqref{eig2clo} in terms of \textit{open topological strings}, thanks to which we can also understand how to perform the analytic continuation
\begin{equation}
Z_{3d/5d}^{(i)}(\xi, q) \widetilde{Z}_{3d/5d}^{(i)}(\xi, \widetilde{q}) \;\; \sim \;\;  
Z_{3d/5d}^{(i)}(\xi, q) \Big/ \widetilde{Z}_{3d/5d}^{(i)}(\xi, \widetilde{q}^{-1})
\end{equation}
In this form the poles cancel almost tautologically, modulo properly taking into account the effect of the $B$-field on the Kahler parameters of the local Calabi-Yau geometry when necessary (for our local $\mathbb{F}_0$ case the $B$-field can be chosen to be zero). We will postpone the open topological string analysis to Section \ref{subs3.6}.

\subsection{Closed bispectral dual Toda (quantized spectral curve)} \label{subs3.4}

Let us now turn to the problem of the quantization of the spectral curve for the closed Toda system in the parameterization \eqref{spe2}, that is
\begin{equation}
\begin{split}
& (i)^{-N} \mathcal{Q}_a(x - i \omega_1) + Q\, (i)^{N} \mathcal{Q}_a(x + i \omega_1) = \left( w - \langle W_{\square}^{SU(2)} \rangle^{NS} + w^{-1} \right) \mathcal{Q}_a(x) \\
& (i)^{-N} \mathcal{Q}_a(x - i \omega_2) + \widetilde{Q}\, (i)^{N} \mathcal{Q}_a(x + i \omega_2) = \left( w - \langle \widetilde{W}_{\square}^{SU(2)} \rangle^{NS} + w^{-1} \right) \mathcal{Q}_a(x)
\end{split}
\end{equation} 
or equivalently
\begin{equation}
\begin{split}
& \left( e^{- i \omega_1 \partial_x} + Q e^{i \omega_1 \partial_x} + e^{2\pi x/\omega_2} + e^{-2\pi x/\omega_2} \right) \mathcal{Q}_a(x) = \langle W_{\square}^{SU(2)} \rangle^{NS} \mathcal{Q}_a(x) \\
& \left( e^{- i \omega_2 \partial_x} + \widetilde{Q} e^{i \omega_2 \partial_x} + e^{2\pi x/\omega_1} + e^{-2\pi x/\omega_1}  \right) \mathcal{Q}_a(x) = \langle \widetilde{W}_{\square}^{SU(2)} \rangle^{NS} \mathcal{Q}_a(x) \label{bisclosed}
\end{split}
\end{equation}
As we already noticed, this is nothing but the bispectral dual system to \eqref{eqfclosed}.
We already saw in Section \ref{subs3.2} that the eigenfunction of the system \eqref{bisclosed} in the $Q=0$ limit is given by the partition function on the squashed three-sphere of a pair of free chiral multiplets (Figure \ref{Fig:simplefree} right):
\begin{equation}
\mathcal{Q}_a(x) = Z_{3d}(x;a) = \mathcal{S}_2(x - a/2 \vert \omega_1, \omega_2) \mathcal{S}_2(x + a/2 \vert \omega_1, \omega_2) \;\;\; \text{at} \;\; Q=0
\end{equation} 
In gauge theory, we can obtain the eigenfunction at $Q \neq 0$ as a series expansion in $Q$ by coupling these two free chiral multiplets (which have a $SU(2)_a \times U(1)_x$ flavour symmetry) to the 5d bulk $SU(2)$ gauge group and compute the 5d instanton partition function $Z_{3d/5d}(x;a)$ in the presence of this new codimension two defect \cite{Gaiotto:2014ina,Bullimore:2014awa}. This is equivalent to weakly gauge the $SU(2)$ flavour symmetry of the two chiral multiplets. Again, this computation can be performed in terms of equivariant characters for vector spaces entering a properly modified ADHM construction, or equivalently in terms of 1d quiver quantum mechanics; we remind to \cite{Gaiotto:2014ina,Bullimore:2014awa} for more details and to Appendix \ref{C} for the formulae we used. The final result reads 
\begin{equation}
Z_{3d/5d}(x;a) = 1 + Q_{5d} \dfrac{q\left[1 + q - qw (\mu^{1/2} + \mu^{-1/2}) \right]}{(1-q)(1-q \mu)(1-q \mu^{-1})(1-w \mu^{1/2})(1-w \mu^{-1/2})} + o(Q_{5d}^2)
\end{equation}
with $w = e^{2\pi x/\omega_2}$ and $x$ equivariant mass parameter for the $U(1)_x$ flavour symmetry. Here we only wrote down the first term in the expansion: higher order terms can be easily computed but are just too complicated to be written down. Following the same arguments we presented already many times in this paper, and with the same caveats, we conclude that the non-perturbatively corrected eigenfunction of the system \eqref{bisclosed} is given by
\begin{equation}
\mathcal{Q}_a(x) = Z_{3d}(x;a) Z_{3d/5d}(x;a) \widetilde{Z}_{3d/5d}(x;a) \label{finalchiral}
\end{equation}
Cancellation of poles will be analysed in Section \ref{subs3.6}. We already discussed in Section \ref{subs3.2} how the $Q=0$ limit of \eqref{finalchiral} reduces to the $Q=0$ limit of \eqref{eig2clo} after a Fourier transformation: in fact this is the expected behaviour for a wave-function under a canonical change of coordinates. At the level of gauge theory, this is the expected action of type IIB $S$ duality. More can and has been done in the literature: in particular, in \cite{Gaiotto:2014ina} the authors were able to show that the $S^4 \times S^1$ index (as well as the ''hemisphere index'') of the 5d $SU(2)$ theory in the presence of two 3d free chiral multiplets reduces to the one of the 5d $SU(2)$ theory in the presence of the 3d (simple) defect given by Figure \ref{Fig:2toda} after a Fourier transformation, which in gauge theory language corresponds to the action of $S$ duality in type IIB strings. 
Although we will not check it explicitly, this motivates us to believe that the same thing will also happen in our squashed $S^5$ case (for $\overline{\omega}_1 = \omega_2$ purely imaginary this is essentially the $S^4 \times S^1$ check of \cite{Gaiotto:2014ina}); this would be in line with the discussion in \cite{Grassi:2013qva}.
Nevertheless, we will have an almost explicit check of this statement in Section \ref{subs3.6}.

\subsection{Comparison with exact spectrum from topological strings} \label{subs3.5}

Let us now stop for a moment and clarify better why in the gauge theory literature on 5d Bethe/Gauge correspondence the need for a non-perturbative completion of the flat space results was not noticed or not felt as important. \\
It is easy to show that the gauge theory results for the eigenvalues of the Toda system coincide with the exact ones obtained in \cite{Grassi:2014zfa,2015arXiv151102860H} ``off-shell'', i.e. before imposing the quantization condition \eqref{BAEexact}. First of all, remember that our eigenvalue is given by the Wilson loop \eqref{W2f} and is well-defined even when $\vert q \vert = 1$, differently from what happens for the eigenfunctions. 
If we expand the Wilson loop \eqref{W2f} around $\mu = e^{2\pi a/\omega_2} \rightarrow \infty$ we remain with
\begin{equation}
\begin{split}
\langle W_{\square}^{SU(2)} \rangle^{NS} & \; = \; \mu^{1/2} + \left(1 + Q_{5d\,} \right) \mu^{-1/2} 
+ Q_{5d\,} \left( q^{-1} + 1 + q \right) \mu^{-3/2} \\
& \; + Q_{5d\,}(1 + Q_{5d\,}) \left( q^{-2} + q^{-1} + 1 + q + q^2 \right) \mu^{-5/2} + o(\mu^{-7/2}) \label{expansion}
\end{split}
\end{equation}
In the open Toda case ($Q_{5d\,} = 0$) this simply reduces to
\begin{equation}
\langle W_{\square}^{SU(2)} \rangle^{NS} \;\; \xrightarrow[Q_{5d\,} \rightarrow 0]{} \;\; \mu^{1/2} + \mu^{-1/2}
\end{equation}
as expected. The expansion \eqref{expansion} can be compared with the energy obtained from topological strings \cite{Grassi:2014zfa,2015arXiv151102860H}. We start by considering the Calabi-Yau mirror to local $\mathbb{F}_0$ which geometrically engineers our 5d $\mathcal{N}=1$ $SU(2)$: this is described by the classical curve 
\begin{equation}
e^x + e^p + z_2 e^{-x} + z_1 e^{-p} = 1 \label{curvemarino}
\end{equation} 
on the $(x,p)$ plane, with $z_1, z_2$ complex structure parameters of the geometry. By redefining coordinates and momenta we arrive at
\begin{equation}
e^p + e^{-p} + e^x + m_0 e^{-x} = E
\end{equation}
with $m_0 = z_2/z_1$ and $E = z_1^{-1/2}$; quantization of this curves leads to \eqref{343} with $m_0 = Q_{5d}$. 
If we now consider the quantum mirror map \cite{Aganagic:2011mi} ($z_1 = z = E^{-2}$, $\,z_2 = m_0 z_1 = m_0 z$ and $q = e^{i \hbar}$)
\begin{equation}
\begin{split}
-t_1 = & \ln z_1 + \widetilde{\Pi}_A(z_1, z_2) =  
\ln z + 2 z (1 + m_0) + z^2 \left(3 + 3 m_0^2 + 2 m_0 (q + 4 + q^{-1}) \right) \\
& + z^3 \left( \frac{20}{3} + \frac{20}{3} m_0^3 + 2(m_0 + m_0^2)(q^2 + 6q + 16 + 6 q^{-1} + q^{-2}) \right) + o(z^4) \\
-t_2 = & \ln z_2 + \widetilde{\Pi}_A(z_1, z_2) = \ln m_0 - t_1
\end{split}
\end{equation}
and we invert it we get, after identifying $t_1$ with the $A$-period (i.e. $Q_1 = e^{-t_1} = e^{-a}$)
\begin{equation}
\begin{split}
E \; = \; \dfrac{1}{\sqrt{z}} & \; = \; Q_1^{-1/2} + \left(1 + m_0 \right) Q_1^{1/2} 
+ m_0 \left( q^{-1} + 1 + q \right) Q_1^{3/2} \\
& \; + m_0 (1 + m_0) \left( q^{-2} + q^{-1} + 1 + q + q^2 \right) Q_1^{5/2} + o(Q_1^{7/2}) \label{entop}
\end{split}
\end{equation}
which reduces to 
\begin{equation}
E \;\; \xrightarrow[m_0 \rightarrow 0]{} \;\; Q_1^{-1/2} + Q_1^{1/2}
\end{equation}
in the open Toda case ($m_0 = 0$). Comparison between \eqref{expansion} and \eqref{entop} is immediate, as expected since the two objects should be the same by definition. 
The ``on-shell'' (numerical) spectrum will therefore also coincide, after the appropriate quantization conditions \eqref{BAEexact} are imposed. Clearly the numerical results will be wrong if one uses \eqref{BAE}, i.e. if one only focuses on one copy of $\mathbb{R}^2_{\epsilon_1} \times \mathbb{R}^2 \times S^1_R$ instead of the whole $S^5$, but since in the gauge theory literature most of the computations are done ``off-shell'' this problem was never really fully appreciated. Of course one could still have noticed something was missing by considering higher dimensional defects (as the eigenfunctions) or the partition function itself (i.e. the quantization conditions), since they are ill-defined for non-generic values of the Omega background parameters; nevertheless these problems do not affect much ``off-shell'' computations in 5d gauge theories if one sticks to generic Omega background.

\subsection{NS open topological strings expansion} \label{subs3.6}

Let us turn again to the quantized spectral curves \eqref{bisclosed}, and let us start by just considering the first one. The associated quantum operator is expected to have the NS limit of the refined open topological string partition function $Z_{\text{top, open}}^{NS}$ as its eigenfunction. The refined open topological string partition function $Z_{\text{top, open}}$ admits a definition as an index, in which case it takes the form
\begin{equation}
Z_{\text{top, open}}(q,t,\mathbf{Q},x) = \text{exp}\left[-F^{\text{BPS}}_{\text{top, open}}(q,t,\mathbf{Q},x) \right]
\end{equation}
with
\begin{equation}
F^{\text{BPS}}_{\text{top, open}}(q,t,\mathbf{Q},x) \;=\; \sum_{n = 1}^{\infty} \sum_{s_1, s_2} 
\sum_{\mathbf{d}} \sum_{R}
D_{R, \mathbf{d}}^{s_1, s_2} \dfrac{q^{n s_1} t^{-n s_2}}{n(1-q^n)} \mathbf{Q}^{n \mathbf{d}} \,\text{Tr}_R e^{2 \pi n \widehat{x} / \omega_2} \label{fully}
\end{equation}
expanded in $\mathbf{Q}$ and $e^{2\pi\widehat{x}/\omega_2}$ small. Here $\mathbf{Q} = (Q_1, \ldots, Q_r) = (e^{-2\pi t_1/\omega_2}, \ldots, e^{-2\pi t_r/\omega_2})$ are the exponentials of the flat closed string moduli $\mathbf{t}$, while $\widehat{x}$ is the flat open string modulus indicating the position of the brane associated to the open strings (and related to the value of a $U(1)$ Wilson line along the boundary of the topological string world-sheet), $q=e^{2\pi R \epsilon_1}$, $t = e^{-2\pi R \epsilon_2}$ and $R$ is a representation of the symmetric group $S_N$ of $N$ elements. In order to understand where this $S_N$ comes from and what the integers $D_{R, \mathbf{d}}^{s_1, s_2}$ count, it is better to review the M-theory construction of open topological strings following \cite{Cecotti:2010fi,Aganagic:2011sg}. \\
We start from M-theory on the geometry 
\begin{equation}
(Y \times TN \times S^1)_{q,t}
\end{equation}
with $Y$ a local Calabi-Yau with isometries $U(1)_1 \times U(1)_2 \times U(1)_R$. The Taub-Nut space $TN$ is non-trivially fibered over $S^1$: going around the circle, the $TN$ coordinates $(z_1, z_2)$ are rotated according to
\begin{equation}
z_1 \rightarrow q z_1 \;\;\;,\;\;\; z_2 \rightarrow t^{-1} z_2
\end{equation}
This corresponds to the action of $U(1)_1 \times U(1)_2$. We can now add $N$ M5 branes wrapping
\begin{equation}
(\mathcal{L} \times \mathbb{C} \times S^1)_{q,t} \label{wrapping}
\end{equation}
with $\mathbb{C}$ the $z_1$ plane inside $TN$ and $\mathcal{L}$ a special Lagrangian 3-cycle in $Y$. The M5 branes partition function on this background is defined as the index ($S_1' = S_1 - S_R$, $S_2' = S_2 - S_R$)
\begin{equation}
\text{Tr}_{\mathcal{H}_{BPS}}(-1)^F q^{S_1'} t^{-S_2'} \mathbf{Q}^H U^R \label{index}
\end{equation}
This index is the one appearing in \eqref{fully}. There are two contributions to this index: the one coming from the light modes on the M5 branes, captured by refined Chern-Simons theory on $\mathcal{L}$, and the one coming from the massive M2 branes BPS states associated to the interaction between the M5 branes. The parameters $\mathbf{Q}$ take into account the bulk M2 brane charges in $H_2(Y, \mathbb{Z})$, while $U$ is the holonomy (in the representation $R$ of $S_N$) of the gauge field on $\mathbb{C} \times S^1$ which arises from the M5 brane two-form $B$ wrapping a non-contractible 1-cycle on $\mathcal{L}$ . To conclude, the integers $D_{R, \mathbf{d}}^{s_1, s_2}$ count the number of BPS states of M2 branes of charge $Q \in H_{2}(Y, \mathbb{Z})$ with spin quantum numbers $s_1, s_2$ and in the representation $R$ of $S_N$. \\

As we said, $Z_{\text{top, open}}^{NS}$ is expected to be the eigenfunction of the first quantum spectral curve in \eqref{bisclosed}. We already know that this will not be enough and that also the action of the second quantum spectral curve in \eqref{bisclosed} has to be taken into account; this implies we will have to multiply the open topological string partition function $Z_{\text{top, open}}^{NS}$ by its tilded version $\widetilde{Z}_{\text{top, open}}^{NS}$. 
We can therefore check if our proposed eigenfunction \eqref{finalchiral} can be put in an open topological string like form, i.e.\footnote{As usual, we can bring the second factor at the denominator by analytic continuation in $\widetilde{q}$.}
\begin{equation}
\mathcal{Q}_a(x) \;\overset{?}{=}\; Z_{\text{top, open}}^{NS}(q,\mathbf{Q},\widehat{x}) \widetilde{Z}_{\text{top, open}}^{NS}(\widetilde{q},\widetilde{\mathbf{Q}},\widehat{x}) \label{acaso}
\end{equation}
once we properly identify the correct flat open modulus $\widehat{x}$. A similar check can be performed for the eigenfunction \eqref{eig2clo} of the ``Fourier transformed'' quantum curve \eqref{eqfclosed} (let us take for definiteness the first one in \eqref{eig2clo}):
\begin{equation}
Z_{3d,1l}^{(1)} Z_{3d/5d}^{(1)}(\xi, a) \widetilde{Z}_{3d,1l}^{(1)} \widetilde{Z}_{3d/5d}^{(1)}(\xi, a) \;\overset{?}{=}\; {Z'}_{\text{top, open}}^{NS}(q,\mathbf{Q},\widehat{x}) \widetilde{Z'}_{\text{top, open}}^{NS}(\widetilde{q},\widetilde{\mathbf{Q}},\widehat{x}) \label{acaso2}
\end{equation}
As we will see, this is indeed the case if we consider the NS limit $t \rightarrow 1$ of the index \eqref{fully}, which in our case reads
\begin{equation}
F^{\text{BPS}}_{\text{top, open}}(q,\mathbf{Q},x) \;=\; \sum_{n = 1}^{\infty} \sum_{s_1} 
\sum_{\mathbf{d}} \sum_{m \in \mathbb{Z}}
D_{m, \mathbf{d}}^{s_1} \dfrac{q^{n s_1}}{n(1-q^n)} \mathbf{Q}^{n \mathbf{d}} e^{2\pi m n \widehat{x}/\omega_2} \label{aux1}
\end{equation}
where
\begin{equation}
D_{m, \mathbf{d}}^{s_1} = \sum_{s_2} D_{m, \mathbf{d}}^{s_1, s_2}
\end{equation}
For convenience, we will split this free energy as
\begin{equation}
F^{\text{BPS}}_{\text{top, open}} \; = \; F^{\text{BPS}}_{+} + F^{\text{BPS}}_{0} + F^{\text{BPS}}_{-} \label{aux2}
\end{equation}
according to $m>0$, $m=0$ or $m<0$. 

\begin{figure}[h]
\centering
\includegraphics[width=0.55\textwidth]{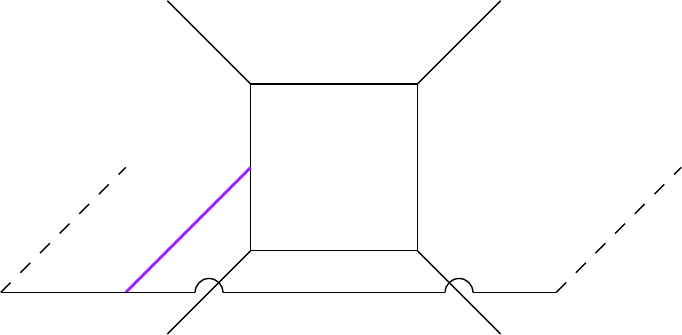}
\caption{Two free chiral multiplets / $D^{(1)}_{0,1}$ defect / type I brane.} \label{Fig:twofree}
\end{figure}

In order to reinterpret our gauge theory computations in terms of topological strings it is more convenient to reduce by a chain of dualities the previously discussed M-theory setting 
to a $(p,q)$-web brane system in the presence of additional D3 branes in type IIB. This is because we know from \cite{Gaiotto:2014ina} how to engineer in type IIB the codimension two vortex defects we used: these come from partial Higgsing of a higher-rank theory \textit{without} defects. For example the codimension two defect associated to the spectral curve \eqref{bisclosed} we studied in Section \ref{subs3.4}, i.e. a pair of chiral multiplets coupled to the $SU(2)$ 5d group, can be obtained from partial Higgsing of the $(p,q)$-web engineering 5d $SU(3)$ $N_F = 2$: partial Higgsing is obtained by fine tuning the mass parameters so that we can move a D5 brane away from the $(p,q)$-web, along a direction transverse to the web. This D5 brane descends from the M5 brane wrapping \eqref{wrapping} in the M-theory construction. By doing so we remain with pure $SU(2)$ with additional D3 branes engineering the defect; see Figure \ref{Fig:twofree}. In \cite{Gaiotto:2014ina} this has been called $D^{(1)}_{0,1}$ defect; it also appeared long before in \cite{Aganagic:2001nx}, where it was named type I brane.

\begin{figure}[h]
\centering
\includegraphics[width=0.25\textwidth]{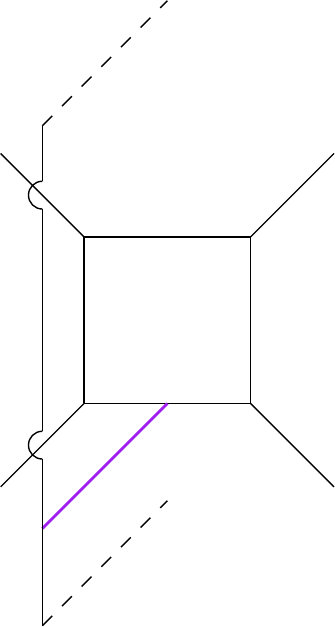}
\caption{$U(1)$ with two chiral multiplets / $D^{(1)}_{1,0}$ defect / type II brane.} \label{Fig:twocoupled}
\end{figure}

Similarly the codimension two defect associated to the spectral curve \eqref{eqfclosed} we studied in Section \ref{subs3.2}, i.e. a $U(1)$ theory with two chiral multiplets coupled to the $SU(2)$ 5d group, can be obtained from partial Higgsing of the $(p,q)$-web engineering 5d $SU(2) \times SU(2)$, this time moving away an NS5 brane from the $(p,q)$-web: see Figure \ref{Fig:twocoupled}. This is known as $D^{(1)}_{1,0}$ defect in \cite{Gaiotto:2014ina} and type II brane in \cite{Aganagic:2001nx}.

It is now clear how to map our gauge theory results to topological strings one. If we denote as $Q_1 = e^{-2\pi t_1/\omega_2}$ the Kahler parameter of the fibre and $Q_2 = e^{-2\pi t_2/\omega_2} = m_0 Q_1$ the Kahler parameter of the base of the local $\mathbb{F}_0$ engineering 5d $SU(2)$, we have the identification
\begin{equation}
Q_1 = e^{-2\pi a/\omega_2} = \mu^{-1} \;\;\;,\;\;\; Q_2 = m_0 Q_1 = Q_{5d} \mu^{-1}
\end{equation}
that is
\begin{equation}
\mu = Q_1^{-1} \;\;\;,\;\;\; Q_{5d} = Q_2 Q_1^{-1}
\end{equation}
The remaining problem would be to properly identify the correct flat open modulus $\widehat{x}$. Luckily, as we already noticed in Section \ref{subs3.5} while comparing eigenvalues, gauge theory quantities are (almost) naturally written in terms of flat open and closed moduli; this is also a consequence of our choice of Baxter-like parameterization for the Seiberg-Witten curve. In fact if we compare with \cite{Aganagic:2001nx} we find that we only have to take into account trivial open mirror maps: for the type $D^{(1)}_{0,1}$ defect this is\footnote{Clearly, according to how one defines spins, the relation between $x$ and $\widehat{x}$ gets modified by $\epsilon_1$ shifts.} 
\begin{equation}
x = \widehat{x} + a/2
\end{equation}
in \eqref{acaso}, while for the type $D^{(1)}_{1,0}$ defect we have
\begin{equation}
-\xi = \widehat{x} + a/2 + i \pi
\end{equation}
in \eqref{acaso2}. If we now focus on the ``holomorphic'' part of the left hand side of \eqref{acaso} and perform an expansion in $Q_1$, $Q_2$ and $e^{2\pi \widehat{x}/\omega_2}$ small we see that we really get an expression of the form \eqref{aux1} which we can later decompose as in \eqref{aux2}; the ``antiholomorphic'' part behaves in the same way. Some of the invariants $D_{m,d_1,d_2}^{s_1}$ for $F_+^{\text{BPS}}$ obtained in this way are listed in Tables \ref{tab3}, \ref{tab4}, \ref{tab88}; $F_-^{\text{BPS}}$ is related to $F_+^{\text{BPS}}$ via
\begin{equation}
F_-^{\text{BPS}}(q, Q_1, Q_2, \widehat{x}) = F_+^{\text{BPS}}(q^{-1}, Q_1, Q_2, -\widehat{x} - a)
\end{equation} 
that is $D_{m,d_1,d_2}^{s_1} = D_{-m,d_1 + m,d_2}^{1-s_1}$ with $m \geqslant 1$, while (modulo constants only depending on $\omega_1$, $\omega_2$)
\begin{equation}
\begin{split}
& F_0^{\text{BPS}}(q, Q_1, Q_2) \;=\; \dfrac{i \pi}{2\omega_1 \omega_2}\left( \widehat{x} + i\dfrac{\omega_1 + \omega_2}{2} \right)^2 + \dfrac{i \pi}{2\omega_1 \omega_2}\left( \widehat{x} + a + i\dfrac{\omega_1 + \omega_2}{2} \right)^2 
- \dfrac{Q_2}{1-q} \\
&\;\;\;\;\; - \dfrac{Q_1 Q_2}{1-q}(q^{-1} + 1) 
- \dfrac{Q_1^2 Q_2}{1-q}(q^{-2} + q^{-1} + 1) 
- \dfrac{Q_2^2}{2(1-q^2)} - \dfrac{Q_1 Q_2^2}{1-q}(2q^{-2} + 2q^{-1} + 1 + q) \\
&\;\;\;\;\; - \dfrac{Q_1^2 Q_2^2}{(1-q^2)}(2q^{-4} + 8q^{-3} + \dfrac{29}{2}q^{-2} + 14q^{-1} + \dfrac{21}{2} + 8q + 6q^2 + 2q^3) + \ldots \label{F0}
\end{split}
\end{equation}
The same thing also happens when we expand the left hand side of \eqref{acaso2}: in this case the invariants $D_{m,d_1,d_2}^{s_1}$ are the same as the \eqref{acaso} ones but with the exchange $Q_1 \leftrightarrow Q_2$ as expected by Fourier transform / type IIB $S$ duality. This provides strong evidence to the claim that \eqref{acaso} and \eqref{acaso2} are related by Fourier transformation.
In the limit $q \rightarrow 1$ these invariants reproduce the ones in \cite{Aganagic:2001nx} for type I and II branes respectively\footnote{It is not yet completely clear to us how to treat type III branes in our language, which most likely will require us to consider dyonic operators as done in \cite{Gang:2012ff} (based on \cite{Drukker:2009id,Gaiotto:2010be,Dimofte:2011jd}) in a similar context.}. For generic $q$ and type $D^{(1)}_{0,1}$ defect we reproduce the invariants given in \cite{Kashani-Poor:2016edc}; the type $D^{(1)}_{1,0}$ defect was not analysed in \cite{Kashani-Poor:2016edc} although it seems likely this can be done with the same techniques discussed in that work. \\
The present discussion motivates the hypothesis that open topological strings as in \eqref{acaso}, \eqref{acaso2} are eigenfuctions of the modular double quantized spectral curves and their bispectral duals. As discussed in \cite{Kashani-Poor:2016edc}, expressions like \eqref{acaso}, \eqref{acaso2} will almost tautologically be free of poles, modulo the possible issue of $B$-field that would shift the Kahler parameters: this is however not a problem in the present example since the $B$-field can be chosen to be zero for local $\mathbb{F}_0$. More in detail, consider for example the logarithm of \eqref{acaso}: this will split into two parts, which we call WKB (``holomorphic'') and non-perturbative (``anti-holomorphic'')
\begin{equation}
\begin{split}
\ln \mathcal{Q}_a(x) \;&=\; \ln Z_{\text{top, open}}^{NS} + \ln \widetilde{Z}_{\text{top, open}}^{NS} \\
& =\; - F^{\text{BPS},\text{WKB}}_{\text{top, open}}(q,\mathbf{Q},\widehat{x}) - F^{\text{BPS},\text{np}}_{\text{top, open}}(\widetilde{q},\widetilde{\mathbf{Q}},\widehat{x})
\end{split}
\end{equation}
This combination reduces to
\begin{equation}
-\sum_{s_1} \sum_{\mathbf{d}} \sum_{m \in \mathbb{Z}} D_{m, \mathbf{d}}^{s_1} \left[
\sum_{n = 1}^{\infty} \dfrac{q^{n s_1}}{n(1-q^n)} \mathbf{Q}^{n \mathbf{d}} e^{2\pi m n \widehat{x}/\omega_2} +
\sum_{\widetilde{n} = 1}^{\infty} \dfrac{\widetilde{q}^{\widetilde{n} s_1}}{\widetilde{n}(1-\widetilde{q}^{\widetilde{n}})} \mathbf{\widetilde{Q}}^{\widetilde{n} \mathbf{d}} e^{2\pi m \widetilde{n} \widehat{x}/\omega_1} \right] \label{pippo1}
\end{equation}
which is more properly rewritten as
\begin{equation}
-\sum_{s_1} \sum_{\mathbf{d}} \sum_{m \in \mathbb{Z}} D_{m, \mathbf{d}}^{s_1} \left[
\sum_{n = 1}^{\infty} \dfrac{q^{n s_1}}{n(1-q^n)} \mathbf{Q}^{n \mathbf{d}} e^{2\pi m n \widehat{x}/\omega_2} -
\sum_{\widetilde{n} = 1}^{\infty} \dfrac{\widetilde{q}^{-\widetilde{n}} \widetilde{q}^{\widetilde{n} s_1}}{\widetilde{n}(1-\widetilde{q}^{-\widetilde{n}})} \mathbf{\widetilde{Q}}^{\widetilde{n} \mathbf{d}} e^{2\pi m \widetilde{n} \widehat{x}/\omega_1} \right] \label{pippo2}
\end{equation}
since $\vert q \vert < 1$ and $\vert \widetilde{q}^{-1} \vert < 1$ when Im$\,(\omega_1/\omega_2) > 0$. The minus sign that appears in going from \eqref{pippo1} to \eqref{pippo2} realizes the analytic continuation in $\widetilde{q}$ we mentioned many times, that is going from expressions like $Z \widetilde{Z}$ written in terms of $q, \widetilde{q}$ to expressions like $Z/\widetilde{Z}'$ written in terms of $q, \widetilde{q}^{-1}$. Notice that the integers $D_{m, \mathbf{d}}^{s_1}$ in the two pieces coincide: this is essential in order to prove poles cancellation. Poles may arise if $\omega_1/\omega_2 = r/s$ with $r,s$ integers when, say, $n = k s$ with $k$ another integer, but they will be cancelled by contributions coming from $\widetilde{n} = k r$. In fact the pole part will be
\begin{equation}
\begin{split}
& D_{m, \mathbf{d}}^{s_1} \left[\dfrac{e^{2\pi i r k s_1}}{2\pi i k^2 s^2 (\omega_1/\omega_2 - r/s)} 
+ \dfrac{e^{2\pi i s k s_1}}{2\pi i k^2 r^2 (\omega_2/\omega_1 - s/r)} \right] e^{-2\pi k \mathbf{d} \cdot \mathbf{t}} e^{2\pi m k \widehat{x}} = \\
& = D_{m, \mathbf{d}}^{s_1} \left[\dfrac{e^{2\pi i r k s_1}}{2\pi i k^2 s^2 (\omega_1/\omega_2 - r/s)} 
- \dfrac{e^{2\pi i s k s_1}}{2\pi i k^2 s^2 (\omega_1/\omega_2 - r/s)} \right] e^{-2\pi k \mathbf{d} \cdot \mathbf{t}} e^{2\pi m k \widehat{x}}
\end{split}
\end{equation}
Poles will therefore cancel when $s_1$ is integer, while if it is half-integer it is necessary to shift the Kahler parameters by a $B$-field; this will be the same shift required in order to cancel poles in the exact quantization condition \eqref{BAEexact}. In our case the spins seem to be all integer, which is consistent with the fact that a $B$-field is not needed for local $\mathbb{F}_0$. \\

To sum up, we have provided some evidence to the validity of the hypothesis that our expressions for the eigenfunctions (left hand side of \eqref{acaso}, \eqref{acaso2}) admit another representation in terms of open topological strings, from which we concluded that our eigenfunctions are regular and free of poles at $\omega_1, \omega_2$ real. At this point, with some expert eye one can see that, were it not for a possible $(-1)^{n(s_1-1)}$ sign missing, \eqref{pippo1} can be completely rewritten in terms of quantum dilogarithms: in fact
\begin{equation}
\begin{split}
& \ln \Big[ \Phi_{\omega_1, \omega_2}\big(m \widehat{x} + i(\omega_1 + \omega_2)(s_1-1/2) - d_1 t_1 - d_2 t_2\big) \Big] = \\
& = - \sum_{n=1}^{\infty} \dfrac{q^n}{n(1-q^n)} q^{n(s_1-1)}(-1)^{n(s_1-1)} 
e^{-2\pi t_1 d_1/\omega_2} e^{-2\pi t_2 d_2/\omega_2} e^{2\pi m n \widehat{x}/\omega_2} \\
& \;\;\; - \sum_{n=1}^{\infty} \dfrac{\widetilde{q}^n}{n(1-\widetilde{q}^n)} \widetilde{q}^{n(s_1-1)}(-1)^{n(s_1-1)} 
e^{-2\pi t_1 d_1/\omega_1} e^{-2\pi t_2 d_2/\omega_1} e^{2\pi m n \widehat{x}/\omega_1}
\end{split}
\end{equation} 
Although this is somehow in contrast with what suggested in \cite{Grassi:2014zfa,2015arXiv151102860H,Kashani-Poor:2016edc}, we propose it is more natural to think of the sign we were missing as the $B$-field contribution, which here appears naturally and restores $\omega_1 \leftrightarrow \omega_2$ symmetry. This might come from a missing sign operator in the definition of index \eqref{index}. We therefore expect our eigenfunctions can be rewritten as
\begin{equation}
\mathcal{Q}_a(x) \propto \mathcal{S}_2(\widehat{x}\vert \omega_1, \omega_2) 
\mathcal{S}_2(\widehat{x} + a\vert \omega_1, \omega_2) 
\sideset{}{'}\prod_{\substack{m \in \mathbb{Z} \\ d_1, d_2, s_1 \\ m \neq 0}}
\left[\Phi_{\omega_1, \omega_2}\Big(m \widehat{x} + i(\omega_1 + \omega_2)(s_1-1/2) - d_1 t_1 - d_2 t_2\Big) \right]^{D_{m,d_1,d_2}^{s_1}} \label{qdr}
\end{equation}
modulo the $\widehat{x}$-independent part of \eqref{F0}. A similar expression will be valid for the Fourier transformed wave-function. Here the primed product means we have to exclude the contributions $(m,d_1,d_2) = (1,0,0)$ and $(m,d_1,d_2) \neq (-1,1,0)$ since they come from the double sine functions; alternatively we can write
\begin{equation}
\begin{split}
\mathcal{Q}_a(x) &\; \propto \; e^{-\frac{i \pi}{2\omega_1 \omega_2}\left( \widehat{x} + i\frac{\omega_1 + \omega_2}{2} \right)^2 + \frac{i \pi}{2\omega_1 \omega_2}\left( \widehat{x} + a + i\frac{\omega_1 + \omega_2}{2} \right)^2 } \times \\
& \times \prod_{\substack{m \in \mathbb{Z} \\ d_1, d_2, s_1 \\ m \neq 0}}
\left[\Phi_{\omega_1, \omega_2}\Big(m \widehat{x} + i(\omega_1 + \omega_2)(s_1-1/2) - d_1 t_1 - d_2 t_2\Big) \right]^{D_{m,d_1,d_2}^{s_1}}
\end{split}
\end{equation}
which more immediately resembles a Chern-Simons theory (light modes on the M5) with massive BPS states corrections (heavy M2 modes). As a side remark, let us point out that (apart from its interpretation in Chern-Simons theories) \eqref{qdr} may have connections to wall crossing along the lines of \cite{Cecotti:2010fi}. \\

Although we did not try to check this because of the computational difficulty of instanton computations, we believe that expanding our eigenfunctions in open topological strings form will produce convergent series in an expansion in $Q_1$, $Q_2$ (instead of the asymptotic $Q_{5d}$ expansion we obtain from gauge theory); this may be also seen from the relation with extended Picard-Fuchs equations (see for example \cite{Fang:2011qd}) or from the quantum dilogarithm representation \eqref{qdr}. Finally, one should be able to reproduce the exact quantization conditions \eqref{BAEexact} by imposing single-valuedness of our proposed eigenfunctions. As far as we understand the techniques for doing so are not yet developed; we refer to \cite{Kashani-Poor:2016edc} for a first step in this direction. \\

\noindent \textit{Comments on fully refined open topological strings} \\

\noindent The integers listed in Tables \ref{tab3}, \ref{tab4}, \ref{tab88} have been computed in \cite{Kashani-Poor:2016edc} for the $D^{(1)}_{0,1}$ defect in a way a bit different from ours. The idea is that, assuming open topological strings are eigenfunctions of the quantum spectral curve, the logarithm of the combination
\begin{equation}
\Xi(x) = \dfrac{\mathcal{Q}_a(x - i \omega_1)}{\mathcal{Q}_a(x)}
\end{equation}
by definition will be of the form
\begin{equation}
\ln \Xi(x) = - \sum_{n = 1}^{\infty} \sum_{s_1} 
\sum_{d_1, d_2} \sum_{m \in \mathbb{Z}}
D_{m, \mathbf{d}}^{s_1} \dfrac{q^{n s_1}}{n(1-q^n)}(q^{-mn} - 1) Q_1^{n d_1} Q_2^{n d_2} e^{2 \pi m n \widehat{x} /\omega_2} \label{fullKPNS}
\end{equation}
This quantity, which is known as $V(X)$ in \cite{Aganagic:2011mi,Grassi:2014zfa}, is actually pretty easy to compute starting from the quantization of the curve \eqref{curvemarino} as an expansion in $z_1$, $z_2$ and then inverting the quantum mirror map. \\
Developing this idea further, one could do more. This same quantity is known as the NS limit of $\mathcal{Y}(x)$ in the $qq$-character literature \cite{Nekrasov:2012xe,Nekrasov:2013xda,Nekrasov:2015wsu,Bourgine:2015szm,Kimura:2015rgi} and can be computed in gauge theory in terms of a series expansion in $Q_{5d}$. Actually one can compute $\mathcal{Y}(x)$ for generic $\epsilon_1$, $\epsilon_2$ \cite{Nekrasov:2015wsu}; this is still expected to be a ratio of the sort 
\begin{equation}
\mathcal{Y}(x;a) = \dfrac{\mathcal{Q'}_a(x - \epsilon_1)}{\mathcal{Q'}_a(x)} \label{chissa}
\end{equation}
Although this more general $\mathcal{Q'}_a$ has not a good definition yet\footnote{Work in progress by N.Nekrasov will clarify the role of this observables in gauge theory.}, $\mathcal{Y}(x)$ has and can be evaluated: see for example \cite{Kim:2016qqs} for more details. Now, since the fully refined open topological string partition function will involve the expression
\begin{equation}
F^{\text{BPS}}_{\text{top, open}}(q,t,\mathbf{Q},x) \;=\; \sum_{n = 1}^{\infty} \sum_{s_1, s_2} 
\sum_{d_1, d_2} \sum_{m \in \mathbb{Z}}
D_{m, d_1, d_2}^{s_1, s_2} \dfrac{q^{n s_1} t^{-n s_2}}{n(1-q^n)} Q_1^{n d_1} Q_2^{n d_2} \,e^{2 \pi mn \widehat{x} / \omega_2} 
\end{equation}
with $q=e^{2\pi R \epsilon_1}$, $t = e^{-2\pi R \epsilon_2}$, if we believe in \eqref{chissa} we could expect that
\begin{equation}
\ln \mathcal{Y}(x;a) = - \sum_{n = 1}^{\infty} \sum_{s_1, s_2} 
\sum_{d_1, d_2} \sum_{m \in \mathbb{Z}}
D_{m, d_1, d_2}^{s_1, s_2} \dfrac{q^{n s_1} t^{-n s_2}}{n(1-q^n)}(q^{-mn} - 1) Q_1^{n d_1} Q_2^{n d_2} e^{2 \pi m n \widehat{x}/\omega_2} \label{fullKP}
\end{equation}
after taking into account the open mirror map. 
We will only take this as an observation, since the relation \eqref{chissa} for $t$ generic is still under study. 
It would be interesting to come back to this point in the future.


\section{Comments} \label{sec4}

Let us sum up the main points discussed in this paper. We saw that eigenfunctions of relativistic quantum mechanical operators such as Toda chains can be thought of as 5d partition functions in the presence of various 3d defect theories; three classes of 3d theories are of special importance to us:
\begin{itemize}
\item Full monodromy defects, related to the eigenfunctions of the Toda Hamiltonians $\widehat{H}_k$ (Figure \ref{Fig:figfullsimple} left);
\item 3d chiral (or anti-chiral) multiplets, related to the eigenfunctions of the quantized spectral curve in the Baxter-like form \eqref{quantumspec} (Figure \ref{Fig:simplefree} right);
\item Simple monodromy defects, $S$-dual to 3d chiral multiplets, related to Fourier transformed eigenfunctions of the quantized spectral curve (Figure \ref{Fig:figfullsimple} right and \ref{Fig:simplefree} left).
\end{itemize}
In the special $SU(2)$ case full and simple monodromy defects coincide. Although formally the 5d partition function on flat space $\mathbb{R}^2_{\epsilon_1} \times \mathbb{R}^2 \times S^1_R$ with a defect living on $\mathbb{R}^2_{\epsilon_1} \times S^1_R$ is an eigenfunction for the corresponding quantum operator when $\hbar$ is complex (in which case the quantum operator is however not self-adjoint and the eigenfunction, as a series in $Q_{5d}$, is not convergent), it has various problems: for example it suffers from quasi-constant ambiguities, and it is ill-defined and presents poles for $\hbar$ real. These problems can be solved if one considers placing the 5d theory on the squashed $S^5_{\omega_1, \omega_2, \omega_3}$ sphere in a particular limit ($\omega_3 \rightarrow 0$); the 3d defect theory will now wrap a squashed three-sphere $S^3_{\omega_1, \omega_2}$ inside $S^5_{\omega_1, \omega_2, 0}$. By doing so we naturally reproduce the modular double structure of the relativistic Toda system proposed in \cite{Kharchev:2001rs}, which seems to be a general property of relativistic quantum operators probably due to their relation to representation theory of quantum groups. The modular double structure doubles the equations our eigenfunctions have to satisfy, thus fixing the quasi-constant ambiguity, and gives exact quantization conditions which take into account the contribution of quantum mechanical instantons; moreover it allows us to consider self-adjoint operators even for $\hbar$ complex. Once written in open topological string form, it is easy to perform analytic continuations in the $\omega_i$'s and see that these non-perturbatively complete eigenfunctions coming from $S^5_{\omega_1, \omega_2, 0}$ no longer have poles for $\hbar$ real once the appropriate $B$-field is taken into account; convergence properties are also improved in this form.  \\
In this paper we used both the gauge theory formalism, in which we know how to treat defects of various codimension
and which are computationally easier since they involve a single expansion parameter $Q_{5d}$ and do not require mirror maps inversion, and the topological string formalism in which all quantities are better defined. It would be interesting to understand if the topological string formalism is enough by itself: however we should first clarify how to compute eigenfunctions in this setting; we hope this work will turn out to be helpful to this scope.
On this point it could be important to remember the Higgsing construction of our 3d defects reviewed in Section \ref{subs3.6}, which is basically telling us that the eigenfunctions of interest (open topological strings) may be recovered from closed topological strings on a bigger local Calabi-Yau three-fold: since \cite{Grassi:2014zfa,2015arXiv151102860H} already provide a good non-perturbative completion of closed topological strings, this idea could be a good direction to follow. A similar idea also appeared in \cite{Hatsuda:2016rmv} in the context of Wilson loop computations for ABJ theory: as far as we understand the claim there is that ABJ Wilson loops, related to open topological strings on local $\mathbb{F}_0$, can be recovered from closed topological strings on the \textit{same} local $\mathbb{F}_0$. It would be very interesting to understand the meaning of this claim on the gauge theory side. \\
Understanding the Fermi gas approach to \cite{Grassi:2014zfa,2015arXiv151102860H} in gauge theory context would also be very interesting. At the moment there is little we can say. The most promising approach would be the free fermion representation of the spectral determinant (as well as of ABJM Wilson loops) \cite{Hatsuda:2013yua,Hatsuda:2016rmv}: the same $\mathfrak{gl}(\infty)$ algebra structure has in fact been noticed in topological strings context \cite{Aganagic:2003qj,Dijkgraaf:2007sw,Aganagic:2011mi} ($I$-brane systems), in gauge theory context \cite{2003hep.th....6238N} (dual instanton partition function), equivariant quantum cohomology of moduli space of torsion free sheaves on $\mathbb{CP}^2$ \cite{2012arXiv1211.1287M} and in many other related settings. \\
A hint on how to proceed may be obtained by considering the four-dimensional limit, in which the same structure also appears and is strictly related to Painlev\'{e} (or $t t^*$) equations. The story goes as follows: Painlev\'{e} equations arise as isomonodromy condition of a system of linear ordinary differential equations or, equivalently, as flatness condition of a particular connection (in gauge theory terms, this is related to the $tt^*$ connection of the Hitchin system associated to a four dimensional theory; see for example \cite{Gamayun:2013auu} for a discussion on the relation between gauge theory and Painlev\'{e} equations). The solution of this system of linear ODE admits an expression as a correlator of a $c=1$ CFT of fermions and ``monodromy fields'' \cite{sato1977} which via Weyl's theorem can be rewritten in terms of Fredholm determinants and Fredholm minors \cite{miwa} (see also \cite{moore1990}).
This has been studied in more detail in \cite{Cecotti:1992vy,Zamolodchikov:1994uw} in the particular case of Painlev\'{e} $PIII_3$, which is related to four dimensional $\mathcal{N}=2$ pure $SU(2)$ theory; these results have later been used in \cite{Bonelli:2016idi} to give a proof of the conjecture of \cite{Grassi:2014zfa} in a particular limit (the ``four dimensional'' limit). Proving the conjecture in full generality would require a modification of this line of reasoning which probably would imply some relation between finite-difference Painlev\'{e} equations and $q-$Virasoro correlators; unfortunately both subjects are not yet well defined nor understood. \\
Finally, it would also be important to have a good non-perturbative completion of gauge theory and topological strings with both $\epsilon_1$ and $\epsilon_2$ turned on: this is required if one wants to study time-dependent relativistic quantum mechanical systems. Clearly the squashed $S^5$ with $\omega_3 > 0$ seems a good direction to follow, but a careful analysis has yet to be performed.

\section*{Acknowledgements}

We thank Giulio Bonelli, and Alessandro Tanzini for suggesting the problem, Hee Joong Chung, Yasuyuki Hatsuda, Amir-Kian Kashani-Poor, Hee-Cheol Kim, Sung-Soo Kim, Peter Koroteev, Kimyeong Lee, Sungjay Lee, Marcos Marino, Fabrizio Nieri, Tomoki Nosaka, Luigi Tizzano, Futoshi Yagi for discussions, comments and correspondence, and especially Alba Grassi for the long discussions and for pointing out many technical problems.

\appendix

\section{Appendix A - Special functions} \label{A}

The quantum dilogarithm $\Phi_{\omega_1, \omega_2}(x)$ admits the integral representation \cite{1995LMaPh..34..249F}
\begin{equation}
\Phi_{\omega_1, \omega_2}(x) = \text{exp} \left( \int_{\mathbb{R} + i 0} \dfrac{e^{-2ixz}}{2\sinh(\omega_1 z)\cdot 2\sinh(\omega_2 z)} \dfrac{dz}{z} \right)
\end{equation}
in the strip $\vert \text{Im}\,z \vert < \vert \text{Im} \left(i \frac{\omega_1 + \omega_2}{2} \right) \vert$. When Im$\,(\omega_1/\omega_2) > 0$ it admits the infinite product representation
\begin{equation}
\Phi_{\omega_1, \omega_2}\left(x + i \frac{\omega_1 + \omega_2}{2}\right) = 
\dfrac{(q e^{2\pi x/\omega_2};q)_{\infty}}{(e^{2\pi x/\omega_1};\widetilde{q}^{-1})_{\infty}} \label{pizza}
\end{equation}
where
\begin{equation}
q = e^{2\pi i \omega_1 /\omega_2} \;\;\;,\;\;\; \widetilde{q} = e^{2\pi i \omega_2 /\omega_1}
\end{equation}
and
\begin{equation}
(w;q)_{\infty} = \prod_{k=0}^{\infty} (1-q^k w)
\end{equation}
The product at the denominator can be brought at the numerator by an analytic continuation in $\widetilde{q}$ according to
\begin{equation}
\prod_{k \geqslant 0} \dfrac{1}{(1-q^{-k}e^{2\pi x/\omega_2})} = \text{exp} \left( \sum_{k \geqslant 1} \dfrac{e^{2\pi k x/\omega_2}}{k(1 - q^{-k})} \right) = \prod_{k \geqslant 0} (1-q^{k+1}e^{2\pi x/\omega_2}) \label{continuation}
\end{equation}
The quantum dilogarithm satisfies the identities
\begin{equation}
\begin{split}
& \Phi_{\omega_1, \omega_2}\left(x + i\frac{\omega_1 + \omega_2}{2} - i \omega_1 \right)
= \left( 1 - e^{2\pi x/\omega_2} \right) \Phi_{\omega_1, \omega_2}\left(x + i\frac{\omega_1 + \omega_2}{2} \right) \\
& \Phi_{\omega_1, \omega_2}\left(x + i\frac{\omega_1 + \omega_2}{2} - i \omega_2 \right)
= \left( 1 - e^{2\pi x/\omega_1} \right) \Phi_{\omega_1, \omega_2}\left(x + i\frac{\omega_1 + \omega_2}{2} \right)
\end{split}
\end{equation}
as well as
\begin{equation}
\Phi_{\omega_1, \omega_2}(x) \Phi_{\omega_1, \omega_2}(-x) = e^{i \pi \frac{x^2}{\omega_1 \omega_2} + i \pi \frac{\omega_1^2 + \omega_2^2}{12\, \omega_1 \omega_2}} 
\end{equation}
We also introduce the double sine function $s_{\omega_1,\omega_2}(x)$, which is defined as the regularization of the infinite product
\begin{equation}
s_{\omega_1,\omega_2}(x) = \prod_{m,n \geqslant 0}\dfrac{m \omega_1 + n \omega_2 + (\omega_1 + \omega_2)/2 - i x}{m \omega_1 + n \omega_2 + (\omega_1 + \omega_2)/2 + i x}
\end{equation}
The double sine is related to the quantum dilogarithm as
\begin{equation}
s_{\omega_1,\omega_2}(x) = e^{-i \pi \frac{x^2}{2\omega_1 \omega_2} - i \pi \frac{\omega_1^2 + \omega_2^2}{24\, \omega_1 \omega_2}} \Phi_{\omega_1, \omega_2}(x)
\end{equation}
and therefore satisfies
\begin{equation}
s_{\omega_1,\omega_2}(x) s_{\omega_1,\omega_2}(-x) = 1
\end{equation}
and admits the infinite product representation
\begin{equation}
s_{\omega_1, \omega_2}\left(x + i \frac{\omega_1 + \omega_2}{2}\right) = 
e^{-i\frac{\pi}{2 \omega_1 \omega_2}\left( x + i (\omega_1 + \omega_2)/2 \right)^2 - \frac{i \pi(\omega_1 + \omega_2)^2}{24 \omega_1 \omega_2} + i \frac{\pi}{12}} \dfrac{(q e^{2\pi x/\omega_2};q)_{\infty}}{(e^{2\pi x/\omega_1};\widetilde{q}^{-1})_{\infty}} \label{product}
\end{equation}
when Im$\,(\omega_1/\omega_2) > 0$. The double sine also satisfies
\begin{equation}
\begin{split}
& s_{\omega_1, \omega_2}\left(x + i\frac{\omega_1 + \omega_2}{2} - i \omega_1 \right)
= i \cdot 2 \sinh\left[\frac{\pi x}{\omega_2}\right] s_{\omega_1, \omega_2}\left(x + i\frac{\omega_1 + \omega_2}{2} \right) \\
& s_{\omega_1, \omega_2}\left(x + i\frac{\omega_1 + \omega_2}{2} + i \omega_1 \right)
= \dfrac{1}{i \cdot 2 \sinh\left[\frac{\pi x}{\omega_2} + i \pi \frac{\omega_1}{\omega_2}\right]} s_{\omega_1, \omega_2}\left(x + i\frac{\omega_1 + \omega_2}{2} \right) \label{property}
\end{split}
\end{equation}
and similar relations with $\omega_1 \leftrightarrow \omega_2$. In the main part of this note we often use the slightly more compact notation
\begin{equation}
\mathcal{S}_2(x \vert \omega_1, \omega_2) = s_{\omega_1, \omega_2}\left(x + i \frac{\omega_1 + \omega_2}{2} \right)
\end{equation}


\section{Appendix B - Tables} \label{B}

In this Appendix we list the integers $D_{m,d_1,d_2}^{s_1}$ discussed in Section \ref{subs3.6}. The number inside the parenthesis indicates the spin $(s_1)$.

\renewcommand\arraystretch{1.4}
\begin{table}[H]
\begin{center}
\begin{tabular}{|c|c|c|c|c|}
\hline   $m=1$   & $d_2 = 0$ &     $d_2 = 1$     &   $d_2 = 2$ & $d_2 = 3$     \\ 
\hline $d_1 = 0$ &   (1) 1   &       (2) 1       & (3) 1       & (4) 1         \\ 
\hline $d_1 = 1$ &           &    (2) 1 (3) 1    & \begin{tabular}{@{}c@{}} (1) 1 (2) 1 (3) 3 \\ (4) 4 (5) 1  \end{tabular} & 
\begin{tabular}{@{}c@{}} (0) 1 (1) 1 (2) 3 \\ (3) 4 (4) 7 (5) 9 \\ (6) 4 (7) 1  \end{tabular} \\ 
\hline $d_1 = 2$ &           & (2) 1 (3) 1 (4) 1 &  \begin{tabular}{@{}c@{}} (0) 2 (1) 4 (2) 4 \\ (3) 8 (4) 11 (5) 11 \\ (6) 4 (7) 1  \end{tabular} & 
\begin{tabular}{@{}c@{}} (-2) 2 (-1) 7 (0) 11 \\ (1) 16 (2) 26 (3) 33 \\ (4) 45 (5) 55 (6) 55 \\ (7) 31 (8) 14 \\ (9) 4 (10) 1  \end{tabular} \\
\hline $d_1 = 3$ &           & \begin{tabular}{@{}c@{}} (2) 1 (3) 1 \\ (4) 1 (5) 1 \end{tabular} & 
\begin{tabular}{@{}c@{}} (-1) 3 (0) 7 (1) 11 \\ (2) 11 (3) 17 (4) 24 \\ (5) 27 (6) 24 (7) 11 \\ (8) 4 (9) 1 \end{tabular}  & \begin{tabular}{@{}c@{}} (-4) 3 (-3) 11 (-2) 31 \\ (-1) 51 (0) 72 (1) 95 \\ (2) 133 (3) 163 (4) 198 \\ (5) 238 (6) 258 (7) 233 \\ (8) 152 (9) 82 (10) 37 \\ (11) 14 (12) 4 (13) 1 \end{tabular}            \\
\hline
\end{tabular} 
\caption{Integers $D_{1,d_1,d_2}^{s_1}$ for the $D^{(1)}_{0,1}$ defect / type I brane. Integers for the $D^{(1)}_{1,0}$ defect / type II brane can be obtained by exchanging $d_1 \leftrightarrow d_2$.} \label{tab3}
\end{center}
\end{table}
\renewcommand\arraystretch{1} 

\renewcommand\arraystretch{1.4}
\begin{table}[H]
\begin{center}
\begin{tabular}{|c|c|c|c|c|}
\hline   $m=2$   & $d_2 = 0$ &     $d_2 = 1$     &   $d_2 = 2$ & $d_2 = 3$  \\ 
\hline $d_1 = 0$ &           &       (3) 1       & (4) 1 (5) 1 & (5) 2 (6) 1 (7) 1 \\ 
\hline $d_1 = 1$ &           &    (3) 1 (4) 1    & 
\begin{tabular}{@{}c@{}} (2) 1 (3) 1 (4) 4 \\ (5) 6 (6) 3 (7) 1 \end{tabular} &
\begin{tabular}{@{}c@{}} (1) 1 (2) 1 (3) 4 \\ (4) 6 (5) 13 (6) 19 \\ (7) 14 (8) 8 \\ (9) 3 (10) 1 \end{tabular}   \\ 
\hline $d_1 = 2$ &           & (3) 1 (4) 1 (5) 1 & 
\begin{tabular}{@{}c@{}} (1) 2 (2) 4 (3) 4 \\ (4) 9 (5) 15 (6) 15 \\ (7) 9 (8) 3 (9) 1 \end{tabular} &
\begin{tabular}{@{}c@{}} (-1) 2 (0) 7 (1) 11 \\ (2) 18 (3) 33 (4) 46 \\ (5) 69 (6) 95 (7) 106 \\ (8) 78 (9) 49 (10) 23 \\ (11) 11 (12) 3 (13) 1 \end{tabular} \\ 
\hline $d_1 = 3$ &           & \begin{tabular}{@{}c@{}} (3) 1 (4) 1 \\ (5) 1 (6) 1 \end{tabular} &
\begin{tabular}{@{}c@{}} (0) 3 (1) 7 (2) 11 \\ (3) 11 (4) 19 (5) 29 \\ (6) 35 (7) 33 (8) 19 \\ (9) 9 (10) 3 (11) 1  \end{tabular} & 
\begin{tabular}{@{}c@{}} (-3) 3 (-2) 11 (-1) 31 \\ (0) 51 (1) 75 (2) 107 \\ (3) 163 (4) 211 (5) 276 \\ (6) 356 (7) 418 (8) 412 \\ (9) 315 (10) 207 (11) 117 \\ (12) 61 (13) 27 (14) 11 \\ (15) 3 (16) 1 \end{tabular} \\
\hline   
\end{tabular} 
\caption{Integers $D_{2,d_1,d_2}^{s_1}$ for the $D^{(1)}_{0,1}$ defect / type I brane. Integers for the $D^{(1)}_{1,0}$ defect / type II brane can be obtained by exchanging $d_1 \leftrightarrow d_2$.} \label{tab4}
\end{center}
\end{table}
\renewcommand\arraystretch{1} 

\renewcommand\arraystretch{1.4}
\begin{table}[H]
\begin{center}
\begin{tabular}{|c|c|c|c|c|}
\hline   $m=3$   & $d_2 = 0$ &     $d_2 = 1$     &     $d_2 = 2$     & $d_2 = 3$  \\ 
\hline $d_1 = 0$ &           &       (4) 1       & (5) 2 (6) 1 (7) 1 & 
\begin{tabular}{@{}c@{}} (6) 3 (7) 3 (8) 3 \\ (9) 1 (10) 1 \end{tabular}  \\ 
\hline $d_1 = 1$ &           &    (4) 1 (5) 1    & 
\begin{tabular}{@{}c@{}} (3) 1 (4) 1 (5) 5 \\ (6) 8 (7) 5 \\ (8) 3 (9) 1 \end{tabular} &
\begin{tabular}{@{}c@{}} (2) 1 (3) 1 (4) 5 \\ (5) 8 (6) 20 (7) 33 \\ (8) 30 (9) 23 (10) 14 \\ (11) 7 (12) 3 (13) 1 \end{tabular}   \\ 
\hline $d_1 = 2$ &           & (4) 1 (5) 1 (6) 1 & 
\begin{tabular}{@{}c@{}} (2) 2 (3) 4 (4) 4 \\ (5) 11 (6) 18 (7) 21 \\ (8) 13 (9) 8 \\ (10) 3 (11) 1 \end{tabular} &
\begin{tabular}{@{}c@{}} (0) 2 (1) 7 (2) 11 \\ (3) 20 (4) 40 (5) 59 \\ (6) 97 (7) 145 (8) 177 \\ (9) 152 (10) 116 (11) 73 \\ (12) 43 (13) 21 (14) 10 \\ (15) 3 (16) 1 \end{tabular} \\ 
\hline $d_1 = 3$ &           & \begin{tabular}{@{}c@{}} (4) 1 (5) 1 \\ (6) 1 (7) 1 \end{tabular} &
\begin{tabular}{@{}c@{}} (1) 3 (2) 7 (3) 11 \\ (4) 11 (5) 21 (6) 34 \\ (7) 43 (8) 43 (9) 28 \\ (10) 17 (11) 8 \\ (12) 3 (13) 1 \end{tabular} & 
\begin{tabular}{@{}c@{}} (-2) 3 (-1) 11 (0) 31 \\ (1) 51 (2) 78 (3) 119 \\ (4) 193 (5) 262 (6) 362 \\ (7) 499 (8) 622 (9) 659 \\ (10) 564 (11) 427 (12) 288 \\ (13) 179 (14) 102 (15) 53 \\ (16) 25 (17) 10 \\ (18) 3 (19) 1  \end{tabular} \\
\hline   
\end{tabular} 
\caption{Integers $D_{3,d_1,d_2}^{s_1}$ for the $D^{(1)}_{0,1}$ defect / type I brane. Integers for the $D^{(1)}_{1,0}$ defect / type II brane can be obtained by exchanging $d_1 \leftrightarrow d_2$.} \label{tab88}
\end{center}
\end{table}
\renewcommand\arraystretch{1} 

\section{Appendix C - Instanton partition functions with defects} \label{C}

As shown in \cite{Gaiotto:2014ina,Bullimore:2014awa}, the instanton part of the partition function for a 5d $\mathcal{N} = 1$ $U(N)$ theory in the presence of a codimension two defect represented by $N$ 3d chiral fields we discussed in Sections \ref{subs3.2} and \ref{subs3.4}, which is $S$-dual to a simple defect, admits the contour integral representation
\begin{equation}
Z_{3d/5d}(x) = \sum_{k = 0}^{\infty} Q_{5d}^k Z_k(x)
\end{equation}
where
\begin{equation}
Z_k(x) = \dfrac{1}{k!} \oint \left[ \prod_{s=1}^k \dfrac{d \left(2\pi R \phi_{s}\right)}{2\pi i} \right] Z_{k}^{5d} Z_k^{3d}(x)
\end{equation}
and
\begin{equation}
\begin{split}
Z_{k}^{5d} = & \left(\dfrac{2\sinh\left[\pi R(\epsilon_1 + \epsilon_2)\right]}{2\sinh\left[\pi R\epsilon_1\right] \cdot 2\sinh\left[\pi R\epsilon_2\right]} \right)^k 
\prod_{\substack{s,t=1 \\ s \neq t}}^k \dfrac{2\sinh\left[\pi R(\phi_s - \phi_t)\right] \cdot 2\sinh\left[\pi R(\phi_s - \phi_t + \epsilon_1 + \epsilon_2)\right]}{2\sinh\left[\pi R(\phi_s - \phi_t + \epsilon_1)\right] \cdot 2\sinh\left[\pi R(\phi_s - \phi_t + \epsilon_2)\right]} \\
& \prod_{s=1}^k \prod_{j=1}^N \dfrac{1}{2\sinh\left[\pi R(\phi_s - a_j + \frac{\epsilon_1 + \epsilon_2}{2})\right] \cdot 2\sinh\left[\pi R(\phi_s - a_j - \frac{\epsilon_1 + \epsilon_2}{2})\right]}
\end{split}
\end{equation}
\begin{equation}
\begin{split}
Z_k^{3d}(x) = \prod_{s=1}^k e^{\pi R\epsilon_2} \dfrac{2\sinh\left[\pi R (\phi_s - x - \frac{\epsilon_1 + \epsilon_2}{2} - \epsilon_2)\right]}{2\sinh\left[\pi R(\phi_s - x - \frac{\epsilon_1 + \epsilon_2}{2})\right]}
\end{split}
\end{equation}
Written in terms of $\sigma_s = e^{2\pi R\phi_s}$, $q_1 = e^{2\pi R \epsilon_1}$, $q_2 = e^{2\pi R \epsilon_2}$, $\mu_j = e^{2\pi R a_j}$, $w = e^{2\pi R x}$ this becomes
\begin{equation}
Z_k(x) = \dfrac{1}{k!} \oint \left[ \prod_{s=1}^k \dfrac{d \sigma_s}{2\pi i \sigma_s} \right] Z_{k}^{5d} Z_k^{3d}(x) \label{fff}
\end{equation}
with
\begin{equation}
\begin{split}
Z_{k}^{5d} = & \left( - \dfrac{1-q_1 q_2}{(1-q_1)(1-q_2)} \right)^k 
\prod_{\substack{s,t=1 \\ s \neq t}}^k \dfrac{(1-\sigma_s \sigma_t^{-1})(1-\sigma_s \sigma_t^{-1} q_1 q_2)}{(1-\sigma_s \sigma_t^{-1} q_1)(1-\sigma_s \sigma_t^{-1} q_2)} \\
& \prod_{s=1}^k \prod_{j=1}^N \dfrac{(- \sqrt{q_1 q_2})}{(1-\sigma_s \mu_j^{-1}\sqrt{q_1 q_2})(1-\sigma_s^{-1}\mu_j \sqrt{q_1 q_2})}
\end{split}
\end{equation}
\begin{equation}
\begin{split}
Z_k^{3d}(x) = \prod_{s=1}^k q_2 \dfrac{1-\sigma_s \,/(w q_2 \sqrt{q_1 q_2})}{1-\sigma_s \,/(w\sqrt{q_1 q_2})}
\end{split}
\end{equation}
The contributing poles correspond to the usual Young tableau coming only from the $Z_{k}^{5d}$ part: for example for $SU(2)$ at lowest $k$ they are
\begin{itemize}
\item $k=1$: poles at $\sigma_1 = \frac{\mu_1}{\sqrt{q_1 q_2}}$ \,/\, $\sigma_1 = \frac{\mu_2}{\sqrt{q_1 q_2}}$ 
\item $k=2$: poles at $\sigma_1 = \frac{\mu_1}{\sqrt{q_1 q_2}}$, $\sigma_2 = \frac{\mu_1}{q_1\sqrt{q_1 q_2}}$ \,/\, $\sigma_1 = \frac{\mu_1}{\sqrt{q_1 q_2}}$, $\sigma_2 = \frac{\mu_1}{q_2\sqrt{q_1 q_2}}$ \,/\, \\
$\sigma_1 = \frac{\mu_2}{\sqrt{q_1 q_2}}$, $\sigma_2 = \frac{\mu_2}{q_1\sqrt{q_1 q_2}}$ \,/\, 
$\sigma_1 = \frac{\mu_2}{\sqrt{q_1 q_2}}$, $\sigma_2 = \frac{\mu_2}{q_2\sqrt{q_1 q_2}}$ \,/\, 
$\sigma_1 = \frac{\mu_1}{\sqrt{q_1 q_2}}$, $\sigma_2 = \frac{\mu_2}{\sqrt{q_1 q_2}}$ \label{jjj}
\end{itemize} 
This coincides with the residue prescription in \cite{Nekrasov:2015wsu} if we reduce to the 4d limit. The 1-loop part will just be given by
\begin{equation}
Z_{3d}(x) = \prod_{j=1}^N \mathcal{S}_2(x - a_j \vert \omega_1, \omega_2)
\end{equation}
with $\sum_{j=1}^N a_j = 0$. These are the formulae that produce expression \eqref{finalchiral} in the main text. \\

One could also consider 3d anti-chiral multiplets, which corresponds to changing $x$ into $-x$; in this case we have to identify $w=e^{-2\pi R x}$, the 1-loop part becomes
\begin{equation}
Z_{3d}(x) = \prod_{j=1}^N \mathcal{S}_2(- x - a_j \vert \omega_1, \omega_2)
\end{equation}
and \eqref{fff} receives other contributions coming from poles of $Z_k^{1d}$ piece, in addition to the poles corresponding to Young tableaux; for $SU(2)$ at lowest $k$ they are
\begin{itemize}
\item $k=1$: $\sigma_1 = \frac{\sqrt{q_1 q_2}}{w}$
\item $k=2$: $\sigma_1 = \frac{\sqrt{q_1 q_2}}{w}$, $\sigma_2 = \frac{\mu_1}{\sqrt{q_1 q_2}}$ \,/\, 
$\sigma_1 = \frac{\sqrt{q_1 q_2}}{w}$, $\sigma_2 = \frac{\mu_2}{\sqrt{q_1 q_2}}$ \,/\, \\
$\sigma_1 = \frac{\sqrt{q_1 q_2}}{w}$, $\sigma_2 = \frac{\sqrt{q_1 q_2}}{q_1 w}$ \,/\, 
$\sigma_1 = \frac{\sqrt{q_1 q_2}}{w}$, $\sigma_2 = \frac{\sqrt{q_1 q_2}}{q_2 w}$
\end{itemize}
This is the residue prescription given in \cite{Gaiotto:2014ina,Bullimore:2014awa}.

\section{Appendix D - Comments on $N$-particle Toda} \label{D}

In this Appendix we will give a few comments on the general relativistic $N$-particle open and closed modular double Toda chain; in particular we want to point out how, in the open Toda case, the eigenfunction of the quantum spectral curve that appears in Sklyanin separation of variables \cite{Kharchev:2001rs} coincides with the one given by gauge theory\footnote{A somewhat similar discussion also appeared in \cite{Bullimore:2014awa}.}. We could choose many different parameterizations for the quantized mirror curves; let us stick for definiteness to the Baxter-like form \eqref{quantumspec}, that is
\begin{equation}
\begin{split}
& (i)^{-N} \mathcal{Q}(x - i \omega_1) + Q\, (i)^{N} \mathcal{Q}(x + i \omega_1) = t_N(w) \mathcal{Q}(x) \\
& (i)^{-N} \mathcal{Q}(x - i \omega_2) + Q\, (i)^{N} \mathcal{Q}(x + i \omega_2) = \widetilde{t}_N(\widetilde{w}) \mathcal{Q}(x) \label{anto}
\end{split}
\end{equation}
with $w = e^{2\pi x/\omega_2}$ and
\begin{equation}
t_N (w) = \sum_{k=0}^N (-1)^k w^{\frac{N}{2} - k} E_k \;\;\;,\;\;\; E_0 = E_N = 1
\end{equation}
The basic idea of separation of variables is to try and construct eigenfunctions of the Toda Hamiltonians $\widehat{H}_k$, $\widehat{\widetilde{H}}_k$ inductively: that is, given $\psi(x_1, \ldots, x_{N-1} \vert \gamma_1, \ldots, \gamma_{N-1})$ satisfying
\begin{equation}
\widehat{t}_{N-1}(w)\, \psi(x_1, \ldots, x_{N-1} \vert \gamma_1, \ldots, \gamma_{N-1}) = t_{N-1}(w)\, \psi(x_1, \ldots, x_{N-1} \vert \gamma_1, \ldots, \gamma_{N-1})
\end{equation}
we want to construct $\psi(x_1, \ldots, x_{N} \vert \gamma_1, \ldots, \gamma_{N})$ satisfying
\begin{equation}
\widehat{t}_{N}(w)\, \psi(x_1, \ldots, x_{N} \vert \gamma_1, \ldots, \gamma_{N}) = t_{N}(w)\, \psi(x_1, \ldots, x_{N} \vert \gamma_1, \ldots, \gamma_{N})
\end{equation}
The notation is the following: $\psi(x_1, \ldots, x_{N} \vert \gamma_1, \ldots, \gamma_{N})$ depends on $N$ coordinate variables $(x_1, \ldots, x_N)$ and $N$ auxiliary variables $(\gamma_1, \ldots, \gamma_N)$ such that $\gamma_1 + \ldots + \gamma_N = 0$ when we decouple the center of mass; the eigenvalues in this case will be $E_1 = \mu_1 + \ldots + \mu_N$ with $\mu_i = e^{2\pi \gamma_i/\omega_2}$, and similarly for $\widetilde{E}_1$ and higher order energies. The authors of \cite{Kharchev:2001rs} showed that if one starts with the solution $\psi(x_1, \ldots, x_{N-1} \vert \gamma_1, \ldots, \gamma_{N-1})$ of the $N-1$ particle open Toda chain and defines
\begin{equation}
\begin{split}
& \Psi_{a_1 + \ldots + a_N}(x_1, \ldots, x_{N} \vert \gamma_1, \ldots, \gamma_{N-1}) = \\
& \text{exp} \left[ \dfrac{i \pi}{\omega_1 \omega_2} \sum_{s=1}^{N-1} \gamma_s^2 - \dfrac{i \pi}{\omega_1 \omega_2} \left( \sum_{j=1}^{N} a_j \right) \left( \sum_{s=1}^{N-1} \gamma_s \right) + \dfrac{2\pi i}{\omega_1 \omega_2}
\left( \sum_{j=1}^{N} a_j - \sum_{s=1}^{N-1} \gamma_s \right) x_N \right] \times \\
& \times \psi(x_1, \ldots, x_{N-1} \vert \gamma_1, \ldots, \gamma_{N-1})
\end{split}
\end{equation}
then the multiple contour integral function
\begin{equation}
\psi(x_1, \ldots, x_{N} \vert a_1, \ldots, a_{N}) = \int \mu(\vec{\gamma}) \mathcal{Q}_{\mathbf{a}}(\vec{\gamma})
\Psi_{a_1 + \ldots + a_N}(x_1, \ldots, x_{N} \vert \gamma_1, \ldots, \gamma_{N-1}) d\vec{\gamma}
\end{equation}
with 
\begin{equation}
\mu(\vec{\gamma}) = \prod_{s<t = 1}^{N-1} 2 \sinh \left[\frac{\pi}{\omega_1} (\gamma_s - \gamma_t) \right] 
\cdot 2 \sinh \left[\frac{\pi}{\omega_2} (\gamma_s - \gamma_t) \right] 
\end{equation}
is an eigenfunction of the $N$ particle open Toda chain if the unknown integral kernel $\mathcal{Q}_{\mathbf{a}}(\vec{\gamma})$ satisfies Baxter's equations
\begin{equation}
\begin{split}
& (i)^{-N} \mathcal{Q}_{\mathbf{a}}(\vec{\gamma} - i \omega_1 \mathbf{e}_s) = \prod_{j=1}^{N} 2 \sinh \left[\frac{\pi}{\omega_2} (\gamma_s - a_j) \right]  \mathcal{Q}_{\mathbf{a}}(\vec{\gamma}) \\
& (i)^{-N} \mathcal{Q}_{\mathbf{a}}(\vec{\gamma} - i \omega_2 \mathbf{e}_s) = \prod_{j=1}^{N} 2 \sinh \left[\frac{\pi}{\omega_1} (\gamma_s - a_j) \right]  \mathcal{Q}_{\mathbf{a}}(\vec{\gamma})
\end{split}
\end{equation}
with $\{ \mathbf{e}_s \}$ standard basis in $\mathbb{R}^N$. These equations can clearly be factorized and reduce to \eqref{anto} in the limit $Q=0$ for each $\gamma_s$; the solution for each factor will be given by $N$ copies of the double sine function. In this way one can reconstruct the $N$ particle eigenfunction $\psi(x_1, \ldots, x_{N} \vert \gamma_1, \ldots, \gamma_{N})$ starting from the 1 particle one $\psi(x_1 \vert \gamma_1) = e^{2\pi i \gamma_1 x_1/(\omega_1 \omega_2)}$. \\
Separation of variables is very natural in terms of gauge theories on $S^3_{\omega_1, \omega_2}$. We know that the simultaneous eigenfunctions of the open Toda Hamiltonians are given by the partition function on $S^3_{\omega_1, \omega_2}$ of the full monodromy defect theory (Figure \ref{Fig:figfullsimple} left). Induction procedure from $N-1$ to $N$ particles corresponds to promote the $U(N-1)$ flavour group to a 3d gauge group, couple it to $N-1$ chiral multiplets with flavour symmetry $U(N)$, dress the partition function with the appropriate Chern-Simons, mixed Chern-Simons and Fayet-Iliopoulos terms, and then integrate over the previous flavour masses. The chiral multiplets with $U(N)$ flavour symmetry are exactly the defects giving the eigenfunction of the quantized spectral curve (Figure \ref{Fig:simplefree} right). Notice that their $S$-dual theories (Figure \ref{Fig:simplefree} left) are not always well-defined: these are $U(1)$ theories with a chiral in the fundamental representation of the flavour symmetry $U(N)$, which presents parity anomalies for $N$ odd; most likely this parity anomaly is related to the necessity of introducing a $B$-field in the Kahler parameters when considering the closed Toda theory in order to ensure poles cancellation in the exact quantization conditions. \\

As far as the $N$-particle closed Toda system is concerned, gauge theory tells us that Sklyanin separation of variables has to be completed by the operation of promoting the 3d defect theory $SU(N)$ flavour group to a 5d gauge group. It would be interesting to analyse this point in more detail: separation of variables in the open case allows us to study the $SU(N)$ defect theory by only knowing the result for the Abelian theory and the integration kernel $\mathcal{Q}_{\mathbf{a}}(x)$, so one could try to see if something similar happens also in the closed case and if this can be of some help. On the gauge theory side there is no much conceptual difference between closed $N$-Toda and what we already saw in the 2-particle case: the only problems correspond to having more complicated formulae and being careful with the $B$-field, which as we just discussed we expect to be present for $N$ odd.

\bibliography{bibl}
\bibliographystyle{JHEP}

\end{document}